\documentclass[journal]{IEEEtran}
\usepackage{caption}
\IEEEoverridecommandlockouts
\usepackage{comment}
\usepackage{verbatim}
\usepackage{booktabs}
\usepackage{multirow}  
\usepackage{booktabs}  
\usepackage{graphicx}  
\usepackage{cite}
\usepackage{amsmath}
\usepackage{multirow}
\usepackage{amsmath,amssymb,amsfonts}
\usepackage{algorithmic}
\usepackage{graphicx}
\usepackage{textcomp}
\usepackage{array}
\usepackage{hyperref}
\usepackage[linesnumbered,ruled,vlined]{algorithm2e}
\usepackage{romannum}
\usepackage{xcolor}
\def\BibTeX{{\rm B\kern-.05em{\sc i\kern-.025em b}\kern-.08em
    T\kern-.1667em\lower.7ex\hbox{E}\kern-.125emX}}
\begin{document}
\title{Security Risks in Vision-Based Beam Prediction:\\ From Spatial Proxy Attacks to Feature Refinement


 \thanks{
}
}
\author{Avi Deb~Raha,~\IEEEmembership{Student Member,~IEEE,}
        Kitae~Kim,~\IEEEmembership{}
        Mrityunjoy~Gain,~\IEEEmembership{Student Member,~IEEE,}
        Apurba~Adhikary,~\IEEEmembership{}
        Zhu~Han,~\IEEEmembership{Fellow,~IEEE,}
        Eui-Nam~Huh,~\IEEEmembership{Member,~IEEE,}
        and~Choong Seon~Hong,~\IEEEmembership{Fellow,~IEEE}
        \vspace{-8mm}
\thanks{Avi Deb Raha, Kitae Kim, Apurba Adhikary, Eui-Nam Huh and Choong Seon Hong are with the Department of Computer Science and Engineering, School of Computing,
Kyung Hee University, Yongin 17104, Republic of Korea. (e-mail: avi@khu.ac.kr; glideslope@khu.ac.kr; apurba@khu.ac.kr; johnhuh@khu.ac.kr; cshong@khu.ac.kr).

Mrityunjoy Gain and Qiao Yu are with the Department of Artificial Intelligence, School of Computing, Kyung Hee University, Yongin 17104, Republic of Korea. (e-mail: gain@khu.ac.kr; qiaoyu@khu.ac.kr).

Zhu Han is with the Electrical and Computer Engineering Department,
University of Houston, Houston, TX 77004 (email: hanzhu22@gmail.com).

Corresponding author: Choong Seon Hong (e-mail: cshong@khu.ac.kr).
}}
\maketitle
\vspace{-4mm}
\begin{abstract}
The rapid evolution towards the sixth-generation (6G) networks demands advanced beamforming techniques to address challenges in dynamic, high-mobility scenarios, such as vehicular communications. Vision-based beam prediction utilizing RGB camera images emerges as a promising solution for accurate and responsive beam selection. However, reliance on visual data introduces unique vulnerabilities, particularly susceptibility to adversarial attacks, thus potentially compromising beam accuracy and overall network reliability. In this paper, we conduct the first systematic exploration of adversarial threats specifically targeting vision-based mmWave beam selection systems. 
Traditional white-box attacks are impractical in this context because ground-truth beam indices are inaccessible and spatial dynamics are complex. To address this, we propose a novel black-box adversarial attack strategy,  termed Spatial Proxy Attack (SPA), which leverages spatial correlations between user positions and beam indices to craft effective perturbations without requiring access to model parameters or labels.
To counteract these adversarial vulnerabilities, we formulate an optimization framework aimed at simultaneously enhancing beam selection accuracy under clean conditions and robustness against adversarial perturbations. We introduce a hybrid deep learning architecture integrated with a dedicated Feature Refinement Module (FRM), designed to systematically filter irrelevant, noisy and adversarially perturbed visual features. Evaluations using standard backbone models such as ResNet-50 and MobileNetV2 demonstrate that our proposed method significantly improves performance, achieving up to an +21.07\% gain in Top-K accuracy under clean conditions and a 41.31\% increase in Top-1 adversarial robustness compared to different baseline models.
\end{abstract}


\section{Introduction}
The advancement of wireless communication technologies is steering towards the sixth generation (6G) networks, aiming to support emerging applications such as the metaverse, intelligent transportation systems (ITS), digital twins, edge intelligence, and cloud gaming \cite{qiao2025exploring, nguyen2025contemporary, avi, EfficientDeep}. Central to these advancements is the integration of massive Multiple-Input Multiple-Output (MIMO) systems operating at higher frequency bands, including millimeter-wave (mmWave) and terahertz (THz) spectra. These high-frequency bands offer substantial data rates but necessitate the use of large antenna arrays and highly directional, narrow beams. However, configuring these precise beams introduces significant beam training overhead, particularly challenging in high-mobility scenarios \cite{lider}.

In dynamic environments like vehicular networks, rapid changes in channel conditions and user positions demand frequent beam adjustments, complicating beam management. To mitigate these challenges, researchers have explored advanced techniques such as predictive beamforming and machine learning-based approaches to reduce overhead and enhance responsiveness. For instance, leveraging positional information of vehicles has been proposed to predict optimal beamforming configurations \cite{position}. While location-based beamforming provides a foundational approach, its accuracy in predicting beamforming indices is limited. Incorporating RGB camera sensor data has been suggested to extract additional features, thereby improving beam prediction accuracy and enhancing received signal power.

Several studies have investigated the use of RGB camera images for predicting optimal beamforming vectors in high-mobility environments, including vehicle-to-infrastructure (V2I), vehicle-to-vehicle (V2V), and unmanned aerial vehicle (UAV) communications. Specifically, works such as \cite{firstvision} and \cite{gourango_1} utilize camera data from base stations (BSs) to predict the optimal beam from a predefined beam codebook for V2I communications. The integration of attention modules to focus on target users has been explored to improve beam prediction accuracy \cite{Attention_beam}. In UAV communications, distinct approaches involve mounting cameras either at ground BSs or on UAVs themselves, capturing surrounding images for beam prediction \cite{UAV_beam_1, UAV_beam2}. Additionally, integrating semantic segmentation modules has been proposed to enhance focus on user locations, refining beam prediction accuracy \cite{JSAC_Beam, Avi_Noms, raha2024advancing}. A environmental robust mechanism leveraging RGB images and GPS has been introduced to improve the reliability of beam predictions under diverse conditions \cite{raha2024advancing}.

Despite these advancements, {\em a critical challenge remains: the adversarial robustness of vision-based beamforming systems}. Deep learning models are known to be susceptible to adversarial attacks, where malicious actors introduce subtle perturbations to inputs, leading to incorrect predictions. In the context of mmWave beamforming, adversarial perturbations to visual inputs can mislead algorithms into selecting suboptimal beams, resulting in degraded signal quality and reduced Quality of Service (QoS) for users. This vulnerability is particularly concerning in high-mobility environments, where rapid and accurate beam adjustments are crucial for maintaining reliable communication.

Recent studies have begun to address these security concerns. For instance, \cite{turn0search0} investigates adversarial attacks on deep learning-based mmWave beam prediction models and proposes mitigation methods such as adversarial training and defensive distillation. Another study \cite{turn0search3} explores adversarial machine learning security problems for 6G, focusing on mmWave beam prediction use cases and proposing mitigation methods to enhance model robustness. Additionally, \cite{turn0search5} examines adversarial security mitigations of mmWave beamforming prediction models using defensive distillation and adversarial retraining. These studies highlight the need for developing vision-based beamforming approaches that are resilient against adversarial manipulations, ensuring the reliability and efficiency of 6G networks in high-mobility scenarios.

While these works \cite{ turn0search0,turn0search3, turn0search5} provide insights into the adversarial robustness of deep learning-based mmWave beam prediction models, they do not address the specific vulnerabilities of vision-based beamforming systems. Unlike traditional numerical or channel-state-information (CSI)-based beam prediction models, vision-based approaches rely on RGB camera images for inferring beam indices. However, to date, {\em no prior study has systematically explored adversarial attacks in vision-based beam selection}. This gap in research is critical, as vision-based beamforming models introduce unique security challenges distinct from those seen in purely signal-based methods. Given the widespread deployment of cameras in BSs for beam prediction, an adversarial attack on these vision-based models could lead to degradation in beam selection accuracy, affecting network reliability. Nevertheless, the lack of direct mapping between input images and beam indices makes adversarial attacks on these systems inherently more challenging compared to standard image classification tasks.

One of the fundamental reasons adversarial attacks are difficult in vision-based mmWave beam prediction is the necessity of beam index labels for each image. Unlike conventional image classification tasks, where the ground truth (e.g., a cat or a dog) is visually interpretable, the correct beam index is a function of complex spatial relationships and RF conditions. Even if an attacker gains access to BS camera images, determining the ground-truth beam index for a given scene remains highly non-trivial without knowledge of the actual beam alignment. This lack of explicit label visibility makes it challenging to craft white-box adversarial attacks, where the attacker needs access to both the model parameters and the ground-truth outputs. Therefore, white-box attacks on vision-based beamforming are extreamly hard to perform in real-world scenarios. Instead, adversaries must rely on black-box attack strategies, where only the input camera images are available, and the goal is to perturb the input to mislead the model into selecting incorrect beam indices. In this context, adversarial attacks on vision-based beamforming require a fundamentally different approach compared to standard image-based attacks. While challenging, these attacks remain feasible, and in this work, {\em we investigate the feasibility of black-box adversarial attacks tailored to vision-based mmWave beam selection}.

Motivated by the identified challenges, this work aims to address the following research questions:
\begin{enumerate}
    \item How can an adversary effectively launch black-box adversarial attacks on vision-based mmWave beam prediction models, considering that the attacker only has access to camera images without knowledge of ground truth beam indices or model parameters? 
    \item Given that the fundamental objective of vision-based mmWave beam prediction systems is achieving high clean accuracy, how can we design more generalizable models that consistently maintain superior performance under normal operating conditions?  
    \item How can we systematically enhance the robustness of vision-based beam prediction models, ensuring they retain high prediction accuracy against adversarial perturbations without significantly compromising their performance on clean data? 
    \item What strategies can simultaneously optimize both clean accuracy and adversarial robustness, thus ensuring better performance across both benign and adversarial scenarios?
\end{enumerate}

To answer these key questions, we develop a novel framework that systematically analyzes the security vulnerabilities of vision-based mmWave beam prediction models and proposes a robust solution. Our work introduces a new adversarial attack strategy specifically designed for black-box settings, formulates a robust beam prediction optimization problem, and presents a hybrid deep learning architecture with an integrated feature refinement module (FRM) to counteract adversarial influences.  The following is a summary of the contributions:
\begin{itemize}
\item In this study, we propose a robust and efficient vision-based beam selection system for mmWave communications, designed to provide accurate beam selection under both clean and adversarial conditions. To the best of our knowledge, this work is the first systematic exploration of adversarial attacks specifically targeting vision-based mmWave beam selection systems in a black box setup.
\item We introduce a novel black-box adversarial attack strategy, {SPA (Spatial Proxy Attack)}, tailored for vision-based beam selection. SPA leverages spatial proxies to craft effective perturbations without access to ground-truth beam indices or model parameters. By exploiting inherent spatial relationships between beam directions and user positions, SPA proves both feasible and effective in misleading beam selection models under black-box constraints.

\item To address the vulnerability to adversarial perturbations, we formulate an optimization problem aimed at simultaneously maximizing beam selection accuracy in clean scenarios and robustness under adversarial and noise-induced perturbations. 
\item To address the formulated problem, we propose a hybrid neural network architecture incorporating a lightweight, simple, yet effective FRM. The FRM systematically filters out irrelevant and noisy visual features, significantly enhancing both clean accuracy and robustness. By jointly optimizing FRM with standard backbone models (e.g., ResNet-50), our approach effectively mitigates the impact of adversarial attacks and random noise, achieving substantial improvements in both clean and perturbed scenarios.

\item Extensive empirical evaluations across several real-world scenarios demonstrate the effectiveness of our proposed method. Compared to different baselines, the FRM-enhanced architecture achieves up to {+21.07\% gain in Top-K accuracy under clean conditions}, up to {+47.66\% improvement under random noise}, and up to {+61.04\% robustness enhancement under adversarial attacks}, Which highlights its strong potential under challenging conditions.

\end{itemize}

\begin{figure*}[t]
\centerline{\includegraphics[width=15cm]{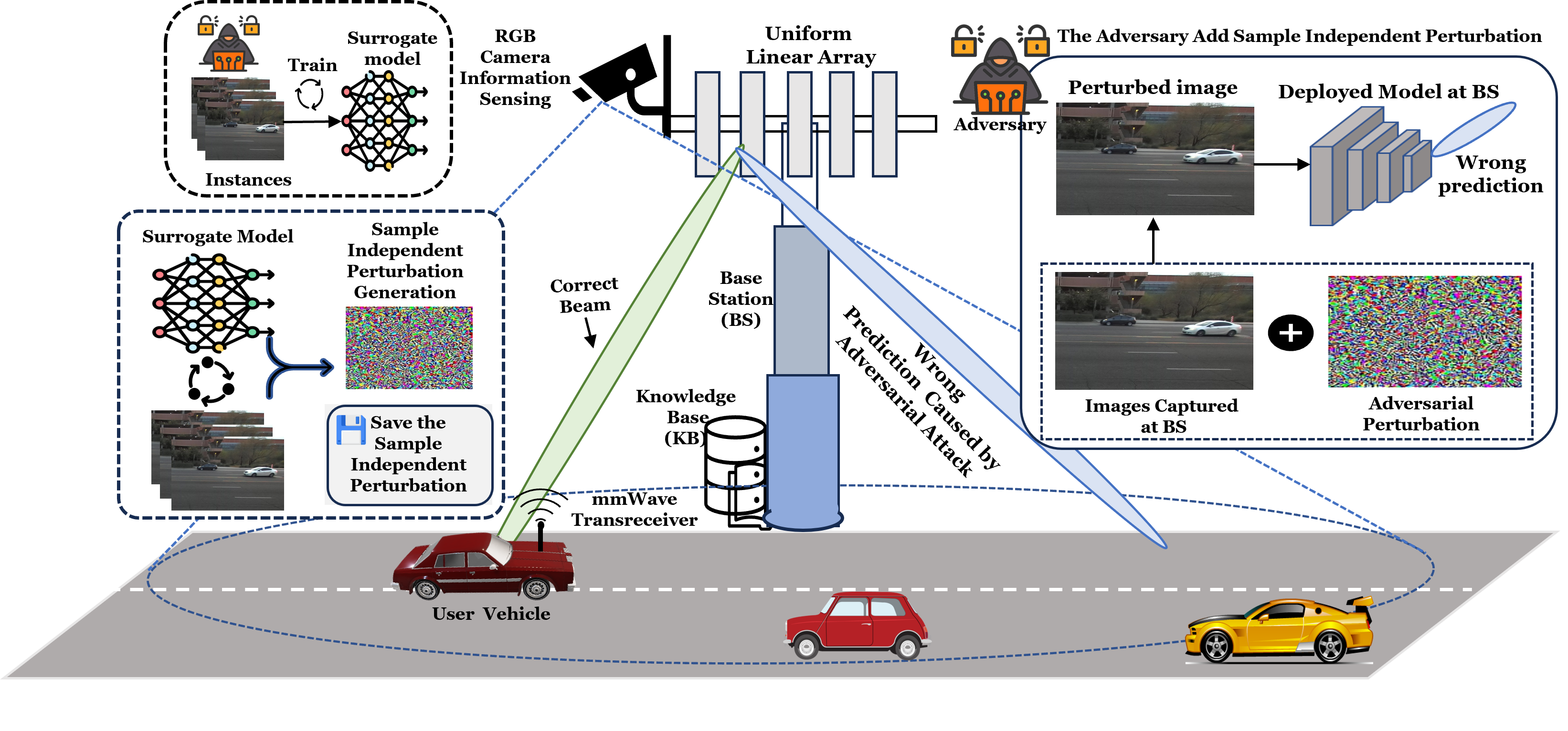}}
\vspace{-5mm}
\caption{System overview of vision-assisted mmWave beam selection with adversarial interference.}
\vspace{-6mm}
\label{System_Model_1}
\end{figure*}
The rest of the paper is organized as follows. Section~\ref{system_model} presents the system model along with the proposed SPA. Section~\ref{problem} outlines the problem formulation. The proposed solution is described in Section~\ref{solution}, and simulation results are analyzed in Section~\ref{results}. Finally, Section~\ref{conclusion} concludes the paper with final remarks.

\section{System Model}\label{system_model}
We consider a BS equipped with an RGB camera, a computational server, a knowledge base (KB), and an extensive array of antennas denoted by \( N_A \gg 1 \). These antennas are arranged in a Uniform Linear Array (ULA) configuration to facilitate precise beamforming. The BS employs an analog beamforming architecture, utilizing a single Radio Frequency (RF) chain. Beamforming vectors are selected from a predefined codebook \( \mathbf{C} \), which is defined as\cite{gourango_1}:
\vspace{-2mm}
\begin{equation}
\mathbf{C} = [\mathbf{c}_1^k, \mathbf{c}_2^k, \dots, \mathbf{c}_i^k, \dots, \mathbf{c}_L^k], \label{eq1}
\vspace{-2mm}
\end{equation}
where \( L \) represents the total number of beamforming vectors, and $k$ represents the number of antenna elements.

The BS is tasked with serving a mobile user \( u_j \) equipped with a single antenna (\( N_R = 1 \)) as illustrated in Fig.~\ref{System_Model_1}. To achieve precise environmental sensing, the RGB camera integrated into the BS captures images \( I[t] \) of the surrounding area at each time slot \( t \). Based on the captured image \( I[t] \) and the predefined beamforming codebook \( \mathbf{C} \), the BS selects the optimal beamforming vector \( \mathbf{c}_i^k[t] \) to accurately serve \( u_j \). This selection process is represented by a function \( f(\cdot) \), which maps the input image \( I[t] \) to a specific beamforming vector from the codebook \( \mathbf{C} \):
\vspace{-2mm}
\begin{equation}
\mathbf{c}_i^k[t] = f(I[t]; \mathbf{C}), \label{eq_function}
\vspace{-2mm}
\end{equation}
where \( f(\cdot) \) denotes the selection function, \( I[t] \in \mathbb{R}^{H \times W \times 3} \) represents the RGB image captured by the BS at time \( t \), with height \( H \), width \( W \), and three color channels, and \( \mathbf{C} \) is the predefined codebook of beamforming vectors. The function \( f(\cdot) \) can either be heuristic or learned (e.g., using a machine learning model), mapping visual features extracted from \( I[t] \) to an appropriate vector in \( \mathbf{C} \).

In this scenario, we consider a block-fading channel model for the downlink communication between the BS and user \( u_j \). The channel vector \( \mathbf{g}_q[t] \) for the \( q^{th} \)-th subcarrier at time \( t \), comprising \( P \) propagation paths, is defined as \cite{lider}:
\vspace{-2mm}
\begin{equation}
\mathbf{g}_q[t] = \sum_{p=1}^{P} \gamma_p \, \boldsymbol{a}(\theta_p^{\text{az}}, \theta_p^{\text{el}}), \label{eq3}
\vspace{-2mm}
\end{equation}
where \( \gamma_p \) represents the complex gain and \( \boldsymbol{a}(\theta_p^{\text{az}}, \theta_p^{\text{el}}) \) denotes the array response vector corresponding to the azimuth angle \( \theta_p^{\text{az}} \) and elevation angle \( \theta_p^{\text{el}} \) of the \( p \)-th path.

The signal received by user \( u_j \) for the \( q^{th} \) subcarrier at time \( t \) is expressed as \cite{RGB_camera}:
\vspace{-2mm}
\begin{equation}
r_q[t] = \mathbf{g}_q^\mathrm{T}[t] \, \mathbf{c}_i^k[t] \, s[t] + n_q[t], \label{eq4}
\vspace{-2mm}
\end{equation}
where \( r_q[t] \in \mathbb{C}^{1 \times 1} \) represents the received signal, \( \mathbf{c}_i[t] \) is the beamforming vector selected for the user \( u_j \) at time \( t \), as determined by the function \( f(\cdot) \), and \( s[t] \in \mathbb{C} \) denotes the transmitted complex symbol at time \( t \), which adheres to the power constraint \( \mathbb{E}[|s[t]|^2] = P_s \). The term \( n_q[t] \) corresponds to the additive white Gaussian noise (AWGN), modeled as a complex Gaussian distribution with zero mean and variance \( \sigma^2 \), i.e., \( n_q[t] \sim \mathcal{CN}(0, \sigma^2) \).

The average received data rate over \( Q \) subcarriers is defined as \cite{semantic_beam}:
\vspace{-2mm}
\begin{equation}
\Gamma[t] = \frac{1}{Q} \sum_{q=1}^{Q} \log_2\left(1 + \frac{P_s}{\sigma^2} \left| \mathbf{g}_q^\mathrm{T}[t] \, \mathbf{c}_i^k[t] \right|^2\right),
\label{eq:nominal_data_rate}
\vspace{-2mm}
\end{equation}
where \( \Gamma[t] \) represents the average data rate at time \( t \). The average data rate $\Gamma$ depends directly on the chosen beamforming vector \(\mathbf{c}_i^k[t]\). In particular, each codebook vector \(\mathbf{c}_i^k \in \mathbf{C}\) yields a different received power \(\left| \mathbf{g}_q^\mathrm{H}[t] \, \mathbf{c}_i^k[t] \right|^2\) and thus affects \(\Gamma[t]\) differently. Consequently, selecting the beam \(\mathbf{c}_i^k[t]\) that maximizes \(\Gamma[t]\) can significantly optimize the data rate. To automate and efficiently implement this selection, we can cast it as a \emph{classification problem}: each beam in the codebook \(\mathbf{C}\) corresponds to a distinct class. Specifically, given the captured image \(I[t]\), the task is to predict the index \(\hat{i}\) of the beam \(\mathbf{c}_{\hat{i}}^k[t]\) that maximizes the expected data rate:
\vspace{-2mm}
\begin{equation}
\hat{i} = \arg\max_{i \in \{1,2,\dots,L\}} \Gamma\bigl(t,\mathbf{c}_i^k\bigr).
\vspace{-2mm}
\end{equation}
This can be expressed with a learnable function \(f_{\theta}(\cdot)\), parameterized by \(\theta\):
\vspace{-2mm}
\begin{equation}
\hat{i} = f_{\theta}(I[t]; \mathbf{C}),
\vspace{-2mm}
\end{equation}
where \(\hat{i}\) is the predicted beam index. In essence, the neural network \(f_{\theta}\) learns to infer which beam is optimal based on features extracted from the image. Once trained, the network enables real-time beam selection by outputting the index \(\hat{i}\) for the codebook vector \(\mathbf{c}_{\hat{i}}^k[t]\).

\subsection{Adversarial Attack Scenario}
\label{sec:adversarial_attack_scenario}
In the proposed vision-assisted beam selection framework, the attacker’s primary objective is to degrade the communication performance of user \( u_j \) by manipulating the BS’s camera feed. To achieve this goal, the attacker introduces carefully crafted perturbations into the camera images, thereby misleading the beam selection function \( f(\cdot) \) into choosing a suboptimal beam. Below, we detail both the attacker’s capabilities and the steps involved in executing such an attack.
\subsubsection{Unauthorized Access to the BS Camera}
We consider a scenario in which an attacker can gain unauthorized access to the RGB camera installed at the BS. Although internal system components, including the beam selection function \( f(\cdot) \), its parameters, and the ground-truth beam indices \(\mathbf{c}_i[t]\) are typically secured within the BS, the camera feed is more susceptible to interception or tampering. We consider, this vulnerability allows the attacker to alter each captured image \(\mathbf{I}[t]\) before it is processed by \( f(\cdot) \).
\subsubsection{Perturbing the Camera Feed to Induce Suboptimal Beam Selection}
At each time slot \( t \), the attacker modifies the camera’s output \(\mathbf{I}[t]\) by adding an imperceptible perturbation \(\boldsymbol{\delta}[t]\in\mathbb{R}^{H\times W\times 3}\):
\begin{equation}
\mathbf{I}'[t] = \mathbf{I}[t] + \boldsymbol{\delta}[t],
\quad \|\boldsymbol{\delta}[t]\|_p \le \epsilon,
\label{eq:perturbed_image}
\end{equation}
where \( p \ge 1 \), \(\epsilon > 0\) is small, and the constraint \(\|\boldsymbol{\delta}[t]\|_p \leq \epsilon\) ensures that the perturbation remains imperceptible to human observers. When the perturbed image \(\mathbf{I}'[t]\) is fed into the beam selection function \( f(\cdot) \), the BS chooses a suboptimal beamforming vector,
\vspace{-2mm}
\begin{equation}
\mathbf{c}_p^k[t] = f\bigl(\mathbf{I}'[t]\bigr),
\quad \text{with}~ \mathbf{c}_k[t] \neq \mathbf{c}_i[t],
\vspace{-2mm}
\label{eq:suboptimal_beam_selection}
\end{equation}
where \(\mathbf{c}_i^k[t]\) is the optimal beam in the absence of malicious interference. This misselection manifests in the received signal at user \(u_j\) , i.e.,
\begin{equation}
r'_q[t] 
= \mathbf{g}_q^\mathrm{H}[t]\,\mathbf{c}_p^k[t]\,s[t] + n_q[t],
\label{eq:received_signal_under_attack}
\end{equation}
resulting in the following degraded average data rate:
\begin{equation}
\Gamma'[t] 
= \frac{1}{Q} \sum_{q=1}^{Q} 
  \log_2\Bigl(1 + \frac{P_s}{\sigma^2}
              \bigl|\mathbf{g}_q^\mathrm{H}[t]\,
              \mathbf{c}_p^k[t]\bigr|^2
        \Bigr).
\label{eq:data_rate_under_attack}
\end{equation}

\subsubsection{Ground-Truth Beam Indices and Surrogate Model Training}
\label{sec:surrogate_training}
To craft effective perturbations, the attacker ideally needs to understand how \(\mathbf{I}[t]\) is mapped to \(\mathbf{c}_i^k[t]\) by the true beam selection function \( f(\cdot) \). In principle, this requires collecting a dataset, i.e.,
\vspace{-2mm}
\[
\mathcal{D} 
= \bigl\{\bigl(\mathbf{I}[t], \mathbf{c}_i^k[t]\bigr) 
   \,\bigm|\, t = 1,2,\dots,n\bigr\},
   \vspace{-2mm}
\]
and then training a \emph{surrogate model} \(\hat{f}(\cdot)\) to approximate \( f(\cdot) \). The surrogate model is optimized using a cross-entropy loss:
\begin{equation}
\mathcal{L}\bigl(\hat{f}\bigr)
= -\frac{1}{n} 
  \sum_{t=1}^{n} 
  \sum_{i=1}^{L}
  \mathbf{y}_i[t] \,\log \hat{p}_i[t],
\label{eq:cross_entropy_loss}
\end{equation}
where \(\mathbf{y}_i[t]\) is the one-hot encoded label indicating the correct beamforming vector \(\mathbf{c}_i^k[t]\), and \(\hat{p}_i[t]\) is the surrogate model’s predicted probability for beam \(i\). 

Although the attacker can access \(\mathbf{I}[t]\) by compromising the camera, the corresponding beam labels \(\mathbf{c}_i^k[t]\) are almost always stored or computed within the BS’s secure environment. This situation renders the direct collection of \(\bigl(\mathbf{I}[t], \mathbf{c}_i^k[t]\bigr)\) pairs infeasible for most real-world attacks. Furthermore, the model parameters \(\theta\) of \( f(\cdot) \) are proprietary and similarly protected.

\subsubsection{Spatial Proxy Attack (SPA)}
\label{sec:novel_attack}
Acquiring ground-truth beam indices \(\mathbf{c}_i^k[t]\) directly from the BS is highly challenging due to stringent security measures; however, our proposed \textit{mechanism} circumvents this requirement by exploiting an inherent property of vision-based beam selection. Specifically, the model \(f_{\theta}(\cdot)\) that maps images \(\mathbf{I}[t]\) to beamforming vectors \(\mathbf{c}_i[t]\) typically relies on spatial cues within the camera feed. Because the BS’s codebook \(\mathbf{C}\) is predetermined and each beam is designed to serve a particular angular or spatial sector, \(\ f_{\theta}(\cdot)\) effectively learns a \emph{location-to-beam} mapping. In other words, if a user is located at a certain position in the camera’s field of view, the system consistently selects (or “points”) a beam that aligns with that position.

Motivated by this, we introduce a \emph{spatial proxy attack} (SPA) approach: rather than trying to mimic the exact beam indices \(\mathbf{c}_i^k[t]\), we segment the image plane horizontally into bins \(\{\chi_1,\ldots,\chi_\mathcal{A}\}\), each covering a distinct horizontal range. For each image \(\mathbf{I}[t]\), we assign a label \(\chi_a\) indicating the bin that contains the user. These bin labels serve as a proxy for the true beam indices because a user’s position in the image is highly correlated with the beam that \(f_{\theta}(\cdot)\) will select. Concretely, if the user appears in bin \(\chi_a\), the system is likely to pick a beam \(\mathbf{c}_i^k[t]\) whose angular orientation matches that region. Hence, misleading the model about the correct bin (i.e., user position) indirectly compels it to choose a suboptimal beam. This proxy-based strategy is not only easier for the attacker, as obtaining bin labels is more straightforward than accessing the BS’s internal beam labels, but also effective, since beam selection in static codebooks is fundamentally driven by user location within the BS’s coverage.

The attacker trains a surrogate model \(\hat{f}_{\text{spatial}}(\cdot)\) to classify \(\mathbf{I}[t]\) into one of the \(\mathcal{A}\) bins, minimizing a cross-entropy loss analogous to \eqref{eq:cross_entropy_loss} but for bin classification. Crucially, causing \(\hat{f}_{\text{spatial}}(\cdot)\) to misclassify the user’s bin is often sufficient to mislead \( f(\cdot) \) into selecting a wrong beam. This bypasses the need for the adversary to know \(\mathbf{c}_i^k[t]\) explicitly. Algorithm~\ref{alg:binning_surrogate_training} outlines the spatial binning process and the training procedure of the surrogate model \(\hat{f}_{\text{spatial}}(\cdot)\) for learning the proxy task used in the attack.

\begin{algorithm}[t]
\caption{Spatial Proxy Labeling and Surrogate Model Training}
\label{alg:binning_surrogate_training}
\begin{algorithmic}[1]
\REQUIRE ~~\\
\textbf{Dataset of Images:} $\mathcal{I} = \{\mathbf{I}[t]\}_{t=1}^{n}$, where each image $\mathbf{I}[t] \in \mathbb{R}^{H\times W\times 3}$\\
\textbf{Object Detector:} $\mathcal{D}(\mathbf{I}[t]) \rightarrow \{(x_{\min}, y_{\min}, x_{\max}, y_{\max})\}$\\
\textbf{Number of Spatial Bins:} $\mathcal{A}$, \textbf{Surrogate Model:} $\hat{f}_{\text{spatial}}(\cdot; \theta)$\\
\textbf{Learning Rate:} $\eta$, \textbf{Epochs:} $E$, \textbf{Loss Function:} Cross-Entropy $\mathcal{L}_{\text{CE}}$

\ENSURE ~~\\
\textbf{Trained Surrogate Model:} $\hat{f}_{\text{spatial}}(\cdot; \theta^{*})$

\vspace{0.5em}
\STATE \textbf{Dataset Labeling:} Initialize dataset $\mathcal{D} = \emptyset$
\FOR{$t=1,\dots,n$}
    \STATE Detect vehicles in $\mathbf{I}[t]$: 
    \vspace{-2mm}
    \[
    \{(x_{\min}^{(k)}, y_{\min}^{(k)}, x_{\max}^{(k)}, y_{\max}^{(k)})\}_k \leftarrow \mathcal{D}(\mathbf{I}[t])
    \vspace{-2mm}
    \]
    \vspace{-2mm}
    \STATE Select one bounding box: $(x_{\min}, y_{\min}, x_{\max}, y_{\max}) \sim \text{Uniform}\left(\left\{(x_{\min}^{(k)}, y_{\min}^{(k)}, x_{\max}^{(k)}, y_{\max}^{(k)})\right\}_{k=1}^{K}\right)$

    \STATE Compute horizontal center:
    \vspace{-2mm}
    \[
    x_c[t] = \frac{x_{\min} + x_{\max}}{2}
    \vspace{-2mm}
    \]
    \vspace{-2mm}
    \STATE Assign spatial bin label $\chi[t] \in \{1,\dots,\mathcal{A}\}$:
    \[
    \chi[t] = \left\lceil\frac{x_c[t]}{W}\cdot\mathcal{A}\right\rceil
    \vspace{-2mm}
    \]
    \STATE Add labeled data: $\mathcal{D} \leftarrow \mathcal{D}\cup \{(\mathbf{I}[t], \chi[t])\}$
\ENDFOR

\vspace{0.5em}
\STATE \textbf{Surrogate Training:} Randomly initialize $\theta$
\FOR{$e=1,\dots,E$ epochs}
    \FOR{each batch $(\mathbf{I}_b,\chi_b)\subseteq\mathcal{D}$}
        \STATE Forward pass:\vspace{-2mm}
        \[
        \hat{\mathbf{p}}(\mathbf{I}_b;\theta) = \text{softmax}\left(\hat{f}_{\text{spatial}}(\mathbf{I}_b;\theta)\right)
        \vspace{-2mm}
        \]
        \vspace{-2mm}
        \STATE Compute cross-entropy loss:
        \[
        \vspace{-2mm}
        \mathcal{L}_{\text{CE}}(\theta) = -\frac{1}{|\mathbf{I}_b|}\sum_{(\mathbf{I},\chi)\in(\mathbf{I}_b,\chi_b)}\log\hat{p}_{\chi}(\mathbf{I};\theta)
        \vspace{-2mm}
        \]
        \STATE Update parameters:
        $\theta\leftarrow\theta - \eta\nabla_{\theta}\mathcal{L}_{\text{CE}}(\theta)$

    \ENDFOR
\ENDFOR

\RETURN Trained surrogate $\hat{f}_{\text{spatial}}(\cdot; \theta^{*})$
\end{algorithmic}
\end{algorithm}

\subsubsection{Sample-Dependent vs. Sample-Independent Attacks}
\label{sec:attack_strategies}
Once the surrogate model \(\hat{f}_{\text{spatial}}(\cdot)\) has been trained, the adversary can generate perturbations in two primary modes:
\begin{algorithm}[h!]
\caption{FGSM-based Universal Adversarial Perturbation (UAP) Generation}
\label{alg:fgsm-uap}
\begin{algorithmic}[1]
\REQUIRE Trained model $f(\cdot;\theta)$, Dataset $\mathcal{D}$, Perturbation bound $\epsilon$, Normalization parameters: mean $\boldsymbol{\mu}$, std $\boldsymbol{\sigma}$
\ENSURE Universal perturbation $\boldsymbol{\delta}_{\text{uni}} \in \mathbb{R}^{3\times H\times W}$
\STATE Set model to evaluation mode and freeze all model parameters.
\STATE Initialize accumulated gradient: $\mathbf{G} \leftarrow \mathbf{0} \in \mathbb{R}^{1 \times 3 \times H \times W}$
\FOR{each batch $\mathbf{I}_b$ in $\mathcal{D}$}
    \STATE Move $\mathbf{I}_b$ to device.
    \STATE Initialize dummy perturbation: $\boldsymbol{\delta} \leftarrow \mathbf{0} \in \mathbb{R}^{1 \times 3 \times H \times W}$ with $\boldsymbol{\delta}$ requiring gradients.
    \STATE Compute perturbed images: $\mathbf{I}_b^{\text{pert}} \leftarrow \Pi_{[0,1]}(\mathbf{I}_b + \boldsymbol{\delta})$
    \STATE Normalize: $\mathbf{I}_b' \leftarrow \frac{\mathbf{I}_b^{\text{pert}} - \boldsymbol{\mu}}{\boldsymbol{\sigma}}$
    \STATE Forward pass: $\mathbf{o} \leftarrow f(\mathbf{I}_b'; \theta)$
    \STATE Compute adversarial loss: $\mathcal{L}_{\text{adv}} \leftarrow -\frac{1}{|\mathbf{I}_b|} \sum_{i=1}^{|\mathbf{I}_b|} \max_{c}\; \mathbf{o}_{i,c}$
    \STATE Backpropagate: compute $\nabla_{\boldsymbol{\delta}} \mathcal{L}_{\text{adv}}$
    \STATE Accumulate gradients: $\mathbf{G} \leftarrow \mathbf{G} + \nabla_{\boldsymbol{\delta}} \mathcal{L}_{\text{adv}}$
\ENDFOR
\STATE Compute average gradient: $\bar{g} \leftarrow \frac{\mathbf{G}}{N} \quad$ (where $N$ is the number of batches)
\STATE Generate universal perturbation: $\boldsymbol{\delta}_{\text{uni}} \leftarrow \epsilon \cdot \text{sign}(\bar{g})$
\STATE Project onto $\ell_\infty$-ball: $\boldsymbol{\delta}_{\text{uni}} \leftarrow \Pi_{\|\cdot\|_{\infty} \leq \epsilon} (\boldsymbol{\delta}_{\text{uni}})$
\RETURN $\boldsymbol{\delta}_{\text{uni}}$
\end{algorithmic}
\end{algorithm}
\paragraph{Sample-Dependent Attack.}
For each incoming image \(\mathbf{I}[t]\), the attacker computes a distinct perturbation \(\boldsymbol{\delta}[t]\) (e.g., via gradient-based optimization) that maximally fools \(\hat{f}_{\text{spatial}}(\cdot)\). Although this per-image approach can achieve high attack success rates, it imposes considerable real-time requirements. Specifically, the attacker would need to:
\begin{enumerate}
    \item Download each fresh image \(\mathbf{I}[t]\) from the BS camera,
    \item Adversarial optimization on \(\hat{f}_{\text{spatial}}(\cdot)\) to obtain \(\boldsymbol{\delta}[t]\),
    \item Re-upload the perturbed image \(\mathbf{I}'[t] = \mathbf{I}[t] + \boldsymbol{\delta}[t]\) to the camera system before the BS processes it.
\end{enumerate}
This cycle introduces significant latency and interaction overhead. Consequently, a sample-dependent attack becomes impractical for real-time scenarios, as each image must undergo a separate optimization process under tight time constraints.

\begin{figure*}[t]
\centerline{\includegraphics[width=17cm]{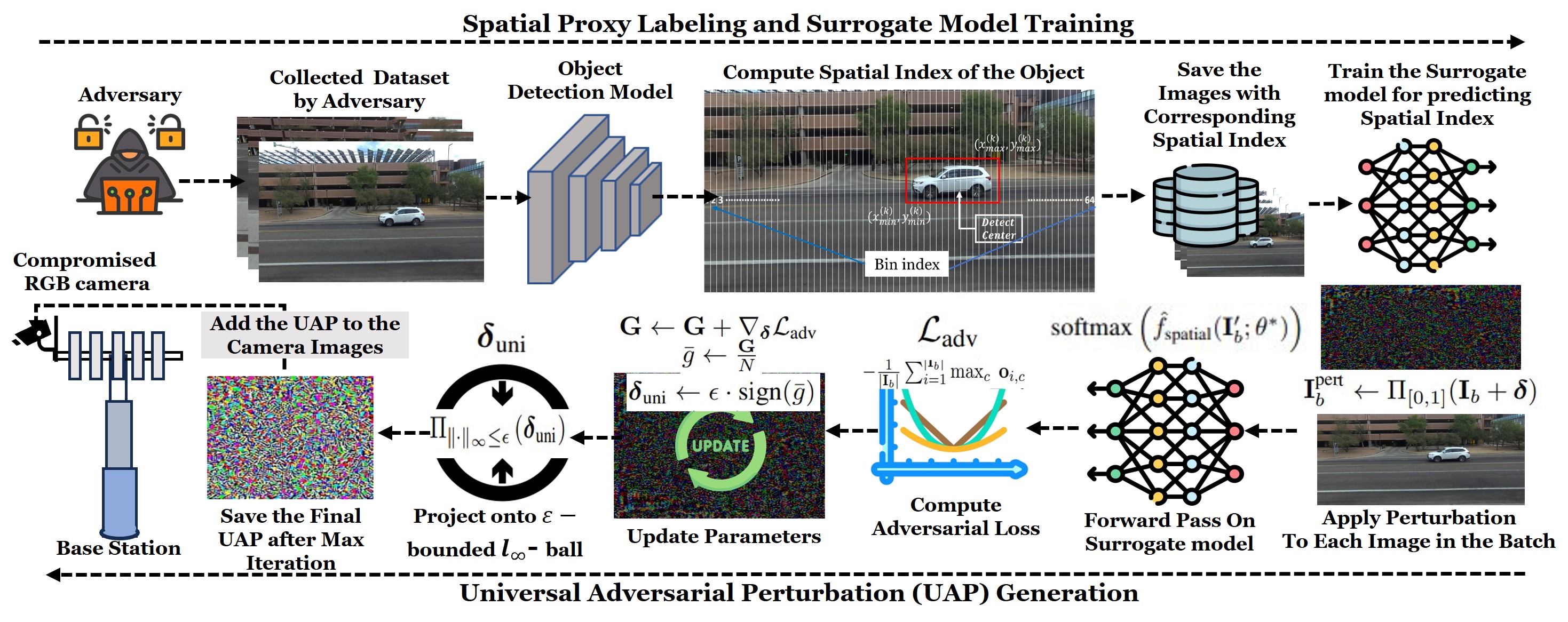}}
\vspace{-2mm}
\caption{Step-by-Step Sample Independent Attack Generation via Spatial Proxy Labeling and Surrogate Model Training.}
\vspace{-5mm}
\label{attack}
\end{figure*}

\paragraph{Sample-Independent Attack.}
In contrast, a sample-independent attack precomputes a single universal perturbation \(\boldsymbol{\delta}_{\text{uni}}\) offline by leveraging the trained surrogate model \(\hat{f}_{\text{spatial}}(\cdot)\) and a representative set of images. The model is trained using spatial-proxy labels obtained by dividing the image plane into \(\mathcal{A}\) bins, each representing a distinct beam sector. To compute \(\boldsymbol{\delta}_{\text{uni}}\), we process each batch \(\mathbf{I}_b\) from the dataset \(\mathcal{D}\) by initializing a dummy perturbation \(\boldsymbol{\delta}\) and applying it to the input:
\vspace{-2mm}
\begin{equation}
\mathbf{I}_b^{\text{pert}} = \Pi_{[0,1]}(\mathbf{I}_b + \boldsymbol{\delta}).
\vspace{-2mm}
\end{equation}
These perturbed images are then normalized using mean \(\boldsymbol{\mu}\) and standard deviation \(\boldsymbol{\sigma}\), followed by a forward pass through the surrogate model:
\vspace{-2mm}
\begin{equation}
\mathbf{o} = \hat{f}_{\text{spatial}}\left( \frac{\mathbf{I}_b^{\text{pert}} - \boldsymbol{\mu}}{\boldsymbol{\sigma}} \right).
\vspace{-2mm}
\end{equation}
An adversarial loss is computed to reduce output confidence:
\vspace{-2mm}
\begin{equation}
\mathcal{L}_{\text{adv}} = -\frac{1}{|\mathbf{I}_b|} \sum_{i=1}^{|\mathbf{I}_b|} \max_{c}\,\mathbf{o}_{i,c}.
\vspace{-2mm}
\end{equation}
Gradients of this loss with respect to \(\boldsymbol{\delta}\) are accumulated over all batches and averaged:
\vspace{-2mm}
\begin{equation}
\bar{g} = \frac{1}{N} \sum_{b=1}^{N} \nabla_{\boldsymbol{\delta}} \mathcal{L}_{\text{adv}}^b.
\vspace{-2mm}
\end{equation}
The final universal perturbation is computed as:
\vspace{-2mm}
\begin{equation}
\boldsymbol{\delta}_{\text{uni}} = \Pi_{\|\cdot\|_{\infty} \leq \epsilon} \left( \epsilon \cdot \text{sign}(\bar{g}) \right).
\vspace{-2mm}
\end{equation}
At test time, this fixed perturbation is added to every incoming image \(\mathbf{I}[t]\), causing consistent misclassification of spatial bins. As a result, the BS is induced to select suboptimal beamforming vectors, enabling an efficient, low-latency black-box attack. Algorithm~\ref{alg:fgsm-uap} illustrates the overall process of generating the sample-independent perturbation. Fig.~\ref{attack} shows the complete workflow of the SPA mechanism.
\vspace{-2mm}
\subsection{Simplistic Noise Attack}
\label{sec:simplistic_noise_attack}
To benchmark the effectiveness of the adversarial perturbations, we also consider a simplistic noise attack. Adding Gaussian noise is straightforward and requires minimal computational resources, making it an attractive simple attack method in many real-world settings. In this scenario, the attacker injects random Gaussian noise $\boldsymbol{\delta}_{\text{noise}}[t]$ into each image:
\begin{equation}
\mathbf{I}''[t] = \mathbf{I}[t] + \boldsymbol{\delta}_{\text{noise}}[t], \quad
\boldsymbol{\delta}_{\text{noise}}[t] \sim \mathcal{N}(0,\sigma_{\text{noise}}^2),
\label{eq:noise_attack}
\end{equation}
where $\sigma_{\text{noise}}^2$ controls the noise magnitude. Though not tailored to the beam selection model, such noise can still degrade image quality and disrupt system performance. Despite its simplicity, this baseline highlights how minimal perturbations can reveal vulnerabilities in vision-based beam selection.

\vspace{-2mm}
\subsection{Attack Pipeline Complexity Analysis}
This section analyzes the computational complexity of Algorithm~\ref{alg:binning_surrogate_training} and Algorithm~\ref{alg:fgsm-uap}, which together constitute the total attack pipeline.
\paragraph{Spatial Proxy Labeling \& Surrogate Training.}
Each of the \(n\) images is labeled via a lightweight object detector, treated as \(\mathcal{O}(1)\) per image due to its fixed model size, yielding a total labeling cost of \(\mathcal{O}(n)\). The surrogate classifier is then trained for \(E\) epochs, each with \(\frac{n}{B}\) batches. If \(F\) denotes the forward/backward pass cost per batch and \(\mathcal{A}\) is the number of bins, each epoch costs \(\mathcal{O}\bigl(\frac{n}{B}\cdot(F + B\cdot \mathcal{A})\bigr)\), simplifying to \(\mathcal{O}(nE)\) when constants are factored out \cite{lecun2015deep}.
\paragraph{Sample-Independent Perturbation Generation.}
Following \cite{goodfellow2015explaining}, the UAP method accumulates gradients over \(\frac{n}{B}\) batches. Each batch incurs a cost of \(\mathcal{O}(F)\) for one forward/backward pass. Hence, the overall complexity is \(\mathcal{O}(n)\) if \(F\) and \(B\) are constants.
\paragraph{Overall Complexity.}
Combining both steps gives \(\mathcal{O}(n) + \mathcal{O}(nE)\), dominated by \(\mathcal{O}(nE)\). In practice, this is considered efficient when the number of epochs \(E\) is relatively small compared to the dataset size \(n\), typically \(E \ll n\), such as \(E \leq 0.01n\). For instance, with \(E = 10\) and typical datasets of size \(n \geq 1000\), the complexity effectively behaves as linear, i.e., \(\mathcal{O}(n)\), making the entire pipeline highly efficient and scalable for real-world applications \cite{goodfellow2016deep}.

\section{Problem Formulation}
\label{problem}
In our system, given a captured image \( \mathbf{I}[t] \) at time \( t \), the beam selection function \( f_{\theta}(\cdot) \) maps the image to a beamforming vector \( \mathbf{c}_i[t] \) from the predefined beamforming codebook \( \mathbf{C} \). However, this mapping is susceptible to disruptions from various sources, including adversarial attacks and random noise. An adversary can deliberately introduce an imperceptible perturbation \( \boldsymbol{\delta}[t] \) to manipulate this mapping, forcing the system to select a suboptimal beam. Additionally, even in the absence of a targeted attack, the presence of random noise in the camera feed can unintentionally degrade the quality of \( \mathbf{I}[t] \), leading to misclassification. 
To mitigate these risks, we aim to optimize the model parameters \( \theta \) such that the beam selection accuracy remains high under three scenarios: clean input, adversarially perturbed input, and noisy input. The problem is formulated as an optimization task where we maximize the beam selection accuracy while ensuring that the model is resilient to both structured adversarial perturbations and simple noise-based distortions. 
The corresponding optimization problem is formulated as follows:
\vspace{-3mm}
\begin{subequations}\label{Opt_1_4}
    \renewcommand{\theequation}{\theparentequation\alph{equation}} 
    \begin{align}
    &P1: \underset{\theta}{\text{maximize}} \;\sum_{t=1}^{T}\;\biggl[ 
      \mathbf{1}\bigl(f_{\theta}(\mathbf{I}[t]) = \mathbf{c}_i[t]\bigr) \notag\\
      &\quad + \max_{\|\boldsymbol{\delta}[t]\|_p \leq \epsilon}
      \mathbf{1}\bigl(f_{\theta}(\mathbf{I}[t]+\boldsymbol{\delta}[t])=\mathbf{c}_i[t]\bigr)\notag\\
      &\quad + \mathbb{E}_{\boldsymbol{\delta}_{\text{noise}}\sim\mathcal{N}(0,\sigma_{\text{noise}}^2)}
      \bigl[\mathbf{1}\bigl(f_{\theta}(\mathbf{I}[t]+\boldsymbol{\delta}_{\text{noise}})=\mathbf{c}_i[t]\bigr)\bigr] 
    \biggr] \tag{19} \label{Opt_1_4:obj}\\[4pt]
    &\text{subject to}\quad \notag\\
    &\quad\mathcal{C}1: \mathbf{c}_i^L \in \mathbf{C}, \quad\forall i, \label{Opt_1_4:const1}\\
    &\quad\mathcal{C}2: \|\boldsymbol{\delta}[t]\|_p \leq \epsilon, \quad\forall t, \label{Opt_1_4:const2}\\
    &\quad\mathcal{C}3: \Gamma[t] \geq \Gamma_{\min}, \quad\forall t, \label{Opt_1_4:const3}\\
    &\quad\mathcal{C}4: \sigma_{\text{noise}}^2 \leq \sigma_{\max}^2. \label{Opt_1_4:const4}
    \end{align}
\end{subequations}
In the propsoed formulation, the first term in the objective function ensures high accuracy in selecting the correct beam under normal conditions. The second term guarantees robustness against worst-case adversarial perturbations within an \( \epsilon \)-bounded norm. The third term accounts for resilience against Gaussian noise by evaluating the expected beam selection accuracy under randomly perturbed images. Unlike adversarial perturbations, which are explicitly optimized to mislead the model, Gaussian noise follows a random distribution and is typically less harmful. However, uncontrolled noise can still degrade performance, making it crucial to assess and mitigate its impact.

Constraint \eqref{Opt_1_4:const1} ensures that the selected beamforming vectors are always within the predefined codebook \( \mathbf{C} \), maintaining compatibility with the system’s physical transmission constraints. Constraint \eqref{Opt_1_4:const2} enforces the adversarial perturbation limit, ensuring that any added perturbation remains within an imperceptible bound \( \epsilon \). Constraint \eqref{Opt_1_4:const3} upholds a minimum data rate threshold \( \Gamma_{\min} \), preventing excessive degradation in communication quality due to either adversarial manipulations or noise-induced errors. Additionally, constraint \eqref{Opt_1_4:const4} imposes an upper bound on the noise variance, \(\sigma_{\text{noise}}^2 \leq \sigma_{\max}^2\), ensuring that the system remains robust under realistic noise levels while avoiding extreme perturbations that would otherwise lead to severe misclassification.

Balancing adversarial robustness, noise resilience, and clean accuracy remains a significant challenge, as conventional adversarial training increases computational overhead and often degrades clean accuracy\cite{QiaoYu_robustness}, while models optimized for clean performance are highly vulnerable to attacks and noise. Although deploying separate models for clean, adversarial, and noisy conditions with dynamic switching appears viable, it is often complex in real-time systems due to the rarity of attacks, overhead from maintaining multiple models, latency introduced by switching, and the difficulty of reliable attack detection. These limitations underscore the need for a unified and efficient approach that ensures robustness without relying on explicit attack detection or model switching, making it more practical for real-time vision-assisted beam selection.
\section{Solution Approach}
\label{solution}
In vision-based beam selection systems, accurate classification fundamentally relies on identifying spatial and positional features directly associated with target users or devices. However, real-world captured images typically include extensive irrelevant information, such as background details, unrelated objects, and environmental textures, which do not contribute meaningfully to the beam selection task\cite{Avi_Noms}. These irrelevant components introduce ambiguity, degrade classification accuracy, and significantly amplify vulnerability to external perturbations. Under clean operating conditions, irrelevant visual details can confuse the model and lead to inaccurate beam predictions\cite{raha2024advancing}. Additionally, in scenarios involving random Gaussian noise, redundant features become channels through which noise distorts critical spatial information, further compromising the model's accuracy. Moreover, adversarial perturbations exploit vulnerabilities by subtly manipulating irrelevant or weakly informative features, inducing misclassifications.

To effectively address these challenges, we propose a dedicated {Feature Refining Module (FRM)} designed to systematically filter irrelevant and noisy features from extracted representations. Unlike traditional adversarial training methods, FRM does not require explicit adversarial examples during training, thus avoiding increased computational complexity and compromised accuracy under normal conditions.
\subsection{Generalized Feature Extraction}
Let the input RGB image at time \( t \) be denoted as \( \mathbf{I}[t]\in \mathbb{R}^{H \times W \times 3} \), where \(H\) and \(W\) represent image height and width. A generic convolutional backbone \(\mathcal{B}_{\theta}(\cdot)\), such as ResNet-50 or MobileNetV2, processes \(\mathbf{I}[t]\) to produce a high-dimensional feature map:
\vspace{-2mm}
\begin{equation}
\mathbf{F}_{\text{backbone}} = \mathcal{B}_{\theta}(\mathbf{I}[t]) \in \mathbb{R}^{C \times H' \times W'},
\vspace{-2mm}
\end{equation}
where \(C\) represents the number of feature channels, and \(H'\), \(W'\) denote spatial dimensions after convolutional processing. Although \(\mathbf{F}_{\text{backbone}}\) contains valuable information about user positions, it may also incorporate redundant or irrelevant visual patterns.
\begin{figure*}[t]
\centerline{\includegraphics[width=16cm]{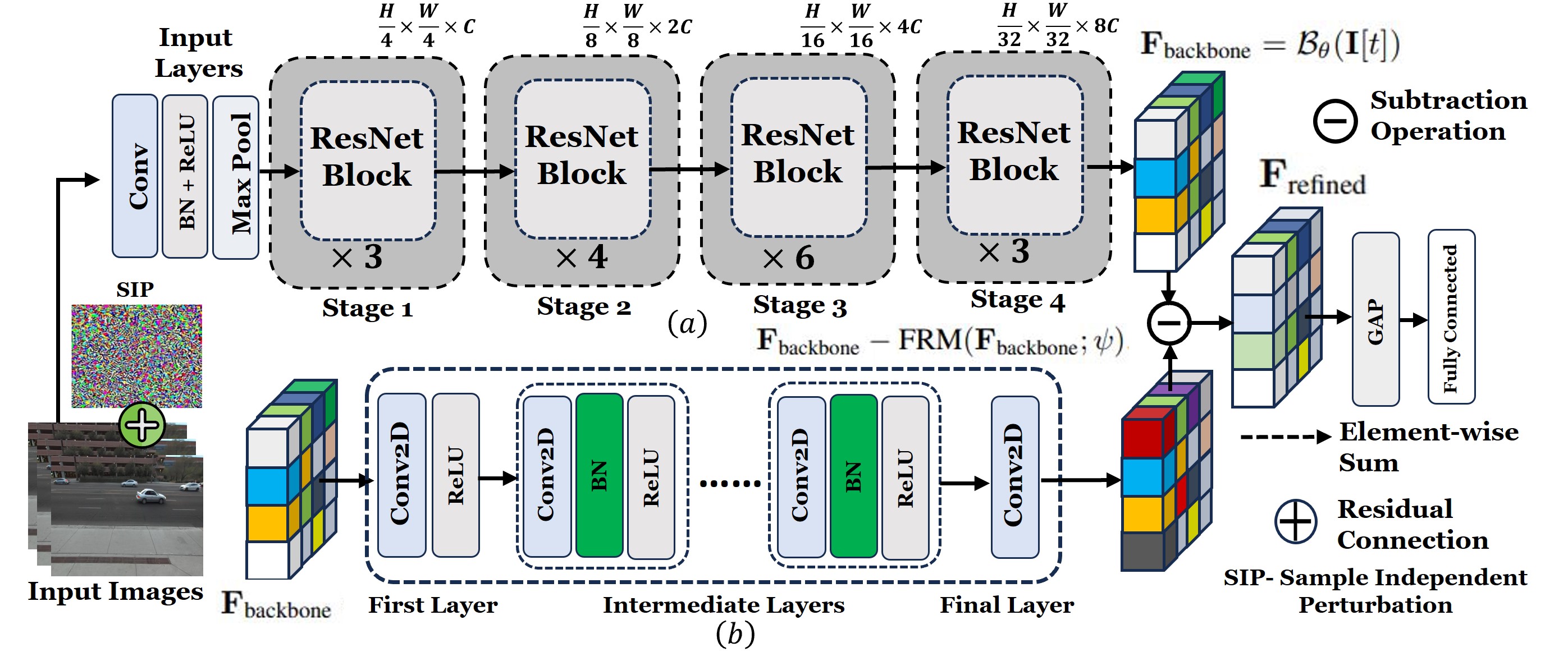}}\vspace{-2mm}
\caption{ Overview of the Proposed FRM-Enhanced ResNet Architecture for Robust Beam Selection.}
\vspace{-3mm}
\label{ResnetFRM}
\end{figure*}
\vspace{-3mm}
\subsection{Feature Refinement via FRM}
To systematically remove unnecessary information, the FRM employs a residual-based refinement strategy. Specifically, the refined feature map \(\mathbf{F}_{\text{refined}}\) is obtained by subtracting the FRM’s learned output from \(\mathbf{F}_{\text{backbone}}\):
\vspace{-2mm}
\begin{equation}
\mathbf{F}_{\text{refined}} = \mathbf{F}_{\text{backbone}} - \text{FRM}(\mathbf{F}_{\text{backbone}}; \psi),
\vspace{-2mm}
\end{equation}
where \(\psi\) denotes parameters of FRM. The module explicitly learns to identify and subtract irrelevant or noisy components within \(\mathbf{F}_{\text{backbone}}\), thereby preserving essential spatial cues.

\subsection{Detailed FRM Architecture}
The FRM consists of sequential convolutional layers designed to progressively refine the feature map:
\begin{enumerate}
\item The first convolutional layer reduces channel dimensions from \(C\) to \(k\):
\vspace{-2mm}
\begin{equation}
\mathbf{F}_{1} = \text{ReLU}(\text{Conv2D}(\mathbf{F}_{\text{backbone}}, \mathbf{W}_1)),
\vspace{-2mm}
\end{equation}
with \(\mathbf{W}_1 \in \mathbb{R}^{3\times3\times C \times k}\).
\item Intermediate layers (layers \(2\) to \(\mathcal{L}-1\)) sequentially apply convolution, batch normalization (BN), and ReLU activation:
\vspace{-2mm}
\begin{equation}
\small
\mathbf{F}_{l} = \text{ReLU}(\text{BN}(\text{Conv2D}(\mathbf{F}_{l-1},\mathbf{W}_{l}))), \quad l=2,...,\mathcal{L}-1.
\vspace{-2mm}
\end{equation}
\item A final convolutional layer restores the feature dimensions to \(C\):
\begin{equation}
\mathbf{F}_{\text{FRM}} = \text{Conv2D}(\mathbf{F}_{\mathcal{L}-1}, \mathbf{W}_\mathcal{L}),
\end{equation}
where \(\mathbf{W}_\mathcal{L} \in \mathbb{R}^{3 \times 3 \times k \times C}\).
\end{enumerate}

The rationale behind the effectiveness of FRM lies in end-to-end optimization. Let  represent the classification loss function (e.g., cross-entropy). The training process jointly optimizes the backbone parameters  and FRM parameters  by minimizing:
\begin{equation}
\small
\min_{\theta,\psi}\sum_{t=1}^{T}\ell\Bigl(f_{\theta}(\mathbf{F}_{\text{backbone}}[t]-\text{FRM}(\mathbf{F}_{\text{backbone}}[t];\psi)), \mathbf{y}_{\text{true}}[t]\Bigr).
\end{equation}

Through this joint optimization, the FRM is explicitly trained to identify and subtract components detrimental to accurate classification, yielding a cleaner representation :
\begin{equation}
\mathbf{F}_{\text{backbone}}[t] = \mathbf{F}_{\text{rel}}[t] + \mathbf{F}_{\text{irr}}[t], \quad \text{FRM}(\mathbf{F}_{\text{backbone}}[t];\psi) \approx \mathbf{F}_{\text{irr}}[t].
\end{equation}
Thus, the refined features primarily contain relevant spatial information necessary for accurate beam selection.

\subsection{Beam Prediction}
The refined feature map \(\mathbf{F}_{\text{refined}}\) is condensed through global average pooling (GAP) into a feature vector:
\vspace{-1mm}
\begin{equation}
\mathbf{z}= \text{GAP}(\mathbf{F}_{\text{refined}}) \in \mathbb{R}^{C},
\vspace{-1mm}
\end{equation}
and passed through a fully connected (FC) layer to produce beam classification scores:
\begin{equation}
\vspace{-1mm}
\mathbf{y} = \text{FC}(\mathbf{z}) \in \mathbb{R}^{L},
\vspace{-1mm}
\end{equation}
where \(L\) denotes the total number of beamforming vectors in the codebook \(\mathbf{C}\). The optimal beam \(\mathbf{c}_i\) is chosen as:
\begin{equation}
\vspace{-1mm}
\mathbf{c}_i = \arg\max(\mathbf{y}).
\vspace{-1mm}
\end{equation}
In summary, the proposed FRM approach provides a unified, computationally efficient solution enhancing clean accuracy, noise resilience, and adversarial robustness, making it particularly suitable for real-time deployment in vision-assisted beam selection systems. Fig. \ref{ResnetFRM} shows the overview of the proposed FRM-enhanced ResNet architecture for robust beam selection.
\vspace{-4mm}

\section{Performance Evaluation}
This section outlines the overall experimental procedure including scenario selection, dataset partitioning, and model setup. It also introduces the baseline methods used for comparison. The following results demonstrate the performance of the proposed attack, along with the clean and robust accuracy of the FRM-enhanced beam prediction framework under multiple real-world conditions.
\subsection{Experimental Setup}
\label{results}
The experimental setup consists of a BS equipped with a 16-element antenna array, operating at the 60 GHz frequency band. The BS is integrated with an RGB camera that captures images at a resolution of $960\times540$, which serve as the sole input for our beam prediction model. The user device ($v_j$) is equipped with a quasi-omni antenna operating in the same band. For all experiments, we use the \emph{DeepSense 6G} dataset~\cite{dataset}, which provides real-world measurements collected from a practical mmWave communication testbed. The dataset includes received power values for 64 predefined transmit beams, enabling us to formulate beam selection as a 64-class classification task. It features a wide range of environments, such as urban intersections, highways, and parking lots, with varying levels of mobility, occlusion, and lighting conditions. 

We focus on Scenarios 1, 2, 3, 4, and 7. Scenarios 1 through 4 involve multiple vehicles in the camera’s field of view, with one vehicle designated as the target. These scenarios test the model’s ability to predict the optimal beam for the target while ignoring irrelevant objects. Scenario 7 includes only a single vehicle and serves as a baseline to evaluate the framework’s performance in a non-interference setting. For training, 50\% of the available data is used. A ResNet-50 architecture is employed as the surrogate model, trained using cross-entropy loss. Vehicle detection is performed using the YOLOv8n model to extract the target vehicle’s bounding box for spatial proxy labeling.

\begin{table*}[htbp]
  \centering
  \caption{Impact of FRM on Clean Accuracy Across Different Scenarios}
  \vspace{-3mm}
  \resizebox{\textwidth}{!}{
    \begin{tabular}{|c|c|c|c|c|c|c|c|}
      \hline
      \multirow{2}{*}{\textbf{Scenario}} & \multirow{2}{*}{\textbf{Top-$k$}} & \multicolumn{3}{c|}{\textbf{ResNet}} & \multicolumn{3}{c|}{\textbf{MobileNet}} \\
      \cline{3-8}
      & & \textbf{FRM+ResNet (\%)} & \textbf{ResNet (\%)} & \textbf{DD Res. (\%)} & \textbf{FRM+MobileNet (\%)} & \textbf{MobileNet (\%)} & \textbf{DD Mob. (\%)} \\
      \hline
      \multirow{4}{*}{Scenario 1} 
        & 1 & \textbf{59.58} & 54.84 & 53.19 & \textbf{58.14} & 54.43 & 53.81 \\ \cline{2-8}
        & 2 & \textbf{83.09} & 80.20 & 79.38 & \textbf{84.32} & 77.93 & 78.76 \\ \cline{2-8}
        & 3 & \textbf{91.54} & 88.45 & 89.89 & \textbf{92.78} & 90.10 & 90.72 \\ \cline{2-8}
        & 5 & \textbf{98.14} & 97.11 & 97.11 & \textbf{98.55} & 96.90 & 97.31 \\ \hline
      \multirow{4}{*}{Scenario 2} 
        & 1 & \textbf{69.39} & 66.05 & 66.72 & \textbf{67.05} & 62.70 & 61.87 \\ \cline{2-8}
        & 2 & \textbf{88.79} & 87.12 & 87.45 & \textbf{86.28} & 85.95 & 82.60 \\ \cline{2-8}
        & 3 & {95.31} & 95.15 & \textbf{95.48} & \textbf{94.98} & 92.97 & 92.64 \\ \cline{2-8}
        & 5 & \textbf{98.99} & 97.65 & 98.49 & \textbf{98.66} & 97.49 & 97.49 \\ \hline
      \multirow{4}{*}{Scenario 3} 
        & 1 & \textbf{55.18} & 51.17 & 51.50 & \textbf{54.51} & 50.50 & 38.79 \\ \cline{2-8}
        & 2 & \textbf{78.92} & 70.56 & 69.89 & \textbf{74.58} & 71.23 & 53.51 \\ \cline{2-8}
        & 3 & \textbf{88.96} & 81.27 & 81.27 & \textbf{84.61} & 82.94 & 66.55 \\ \cline{2-8}
        & 5 & \textbf{96.98} & 92.97 & 90.96 & \textbf{94.98} & 92.97 & 74.24 \\ \hline
      \multirow{4}{*}{Scenario 4} 
        & 1 & \textbf{53.60} & 49.86 & 53.06 & \textbf{50.93} & 45.30 & 42.13 \\ \cline{2-8}
        & 2 & \textbf{78.13} & 74.93 & 77.6 & \textbf{73.86} & 68.80 & 64.8 \\ \cline{2-8}
        & 3 & \textbf{88.80} & 86.93 & 87.2 & \textbf{86.60} & 78.40 & 77.06 \\ \cline{2-8}
        & 5 & \textbf{97.06} & 94.66 & 94.66 & \textbf{96.00} & 90.40 & 87.46 \\ \hline
      \multirow{4}{*}{Scenario 7} 
        & 1 & \textbf{43.60} & 41.86 & 36.62 & \textbf{39.53} & 33.72 & 29.06 \\ \cline{2-8}
        & 2 & \textbf{64.53} & 63.37 & 65.69 & \textbf{61.04} & 52.32 & 49.41 \\ \cline{2-8}
        & 3 & \textbf{79.06} & 73.25 & 75.00 & \textbf{76.16} & 73.25 & 59.30 \\ \cline{2-8}
        & 5 & 87.20 & \textbf{87.79} & \textbf{87.79} & \textbf{87.20} & 84.88 & 76.74 \\ \hline
    \end{tabular}
  }
  \label{tab:clean_accuracy}
  \vspace{-3mm}
\end{table*}

To ensure a balanced and fair evaluation of our proposed framework, we partition the dataset into training, validation, and testing subsets. Specifically, we use {70\%} of the data from each scenario for training, {10\%} for validation to determine the best model checkpoints, and the remaining {20\%} for final performance evaluation. The input RGB images are resized to $224\times224$ for compatibility with the deep neural network architectures. Training is conducted using a mini-batch size of 32 over 30 epochs. The Adam optimizer is employed with an initial learning rate of $10^{-3}$, which is reduced by a factor of 0.1 based on validation performance. The model is trained using the cross-entropy loss function, given the classification nature of the beam selection task. These consistent simulation settings ensure that the results accurately reflect the improvements provided by our proposed approach.
\vspace{-4mm}
\subsection{Baselines}
\vspace{-1mm}
To comprehensively evaluate the robustness of our proposed FRM-enhanced architecture for beam prediction, we benchmark its performance against two well-established convolutional backbones: ResNet-50 and MobileNetV2. The ResNet-50 architecture serves as a primary baseline, selected due to its widespread adoption in previous beam classification studies~\cite{gourango_1, gourango_2}, highlighting its strong representational capability for vision-based beam selection.
Conversely, MobileNetV2 represents a lightweight counterpart, ideal for real-time scenarios demanding frequent and computationally efficient beam alignment\cite{raha2024advancing}. Each backbone is implemented in two distinct configurations: a standard baseline trained using conventional cross-entropy loss, and a defensive distillation (DD) version. The DD variants adopt a student-teacher framework, where the student model learns from both hard ground-truth labels and soft labels generated by a robust teacher model. 

\vspace{-3mm}
\subsection{Numerical Results}
\subsubsection{Impact of FRM on Clean Accuracy}
The clean accuracy improvements attributed to the proposed FRM across various scenarios and top-$k$ beam selection criteria are summarized in Table~\ref{tab:clean_accuracy}. Overall, the incorporation of FRM consistently enhances model accuracy for both ResNet-50 and MobileNetV2, demonstrating the effectiveness of explicit feature refinement even without adversarial training.

In {Scenario 1}, FRM notably increases accuracy at Top-1 for both architectures, with ResNet-50 improving by 4.74\% (54.84\% to 59.58\%) and MobileNetV2 by 3.71\% (54.43\% to 58.14\%). The DD versions underperform slightly compared to the baselines, primarily because DD softens decision boundaries through high-temperature training, which can overly generalize learned representations and reduce accuracy in clean conditions~\cite{papernot2016distillation, carlini2017adversarial}. For Top-5, all models reach high accuracy, but FRM consistently outperforms DD and other base variants. {Scenario 2} presents similar trends, with notable FRM gains at Top-1 (ResNet-50: 3.34\%, MobileNetV2: 4.35\%). DD variants closely follow or slightly surpass baseline accuracy only at higher Top-$k$ predictions, yet fail to match FRM’s consistent improvements. In {Scenario 3}, FRM significantly surpasses both baseline and DD models, especially at Top-2 with ResNet-50 improving by 8.36\% (70.56\% to 78.92\%). The DD versions, particularly MobileNetV2, exhibit substantial accuracy degradation (up to 16.67\%) due to overly softened representations that fail to capture crucial spatial details, making them less effective in complex scenarios.

For {Scenario 4}, MobileNetV2 achieves its largest improvement (8.20\% at Top-3) with FRM, whereas DD models again lag behind due to generalized decision boundaries causing reduced specificity in predictions. Lastly, in {Scenario 7}, DD severely underperforms at Top-1 due to excessive generalization, while FRM still provides substantial improvements (e.g., 5.81\% for MobileNetV2). Overall, the explicit feature refinement of FRM consistently outperforms DD and the base variants by preserving essential spatial details critical for accurate beam classification.

\begin{figure*}[t]
\centerline{\includegraphics[width=18cm]{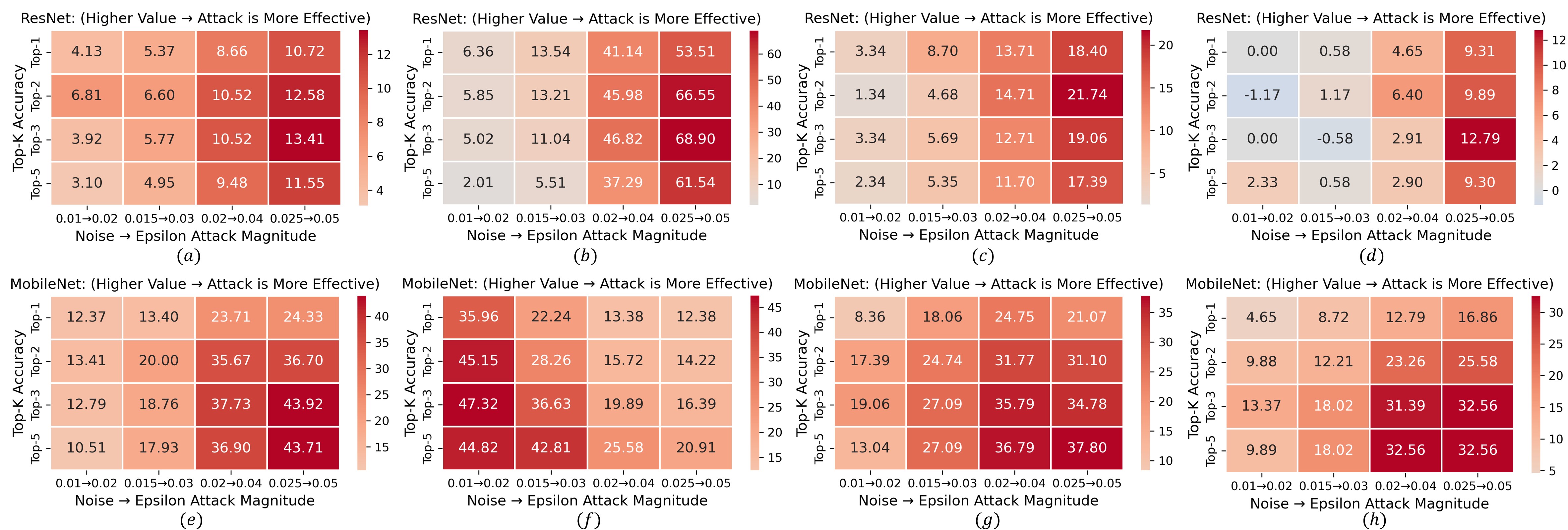}}
\vspace{-2mm}
\caption{Comparison of the impact of Gaussian noise vs. adversarial attacks. The heatmaps depict Top-K accuracy degradation difference (Noise - Adversarial) for ResNet-50 (top) and MobileNetV2 (bottom) across four scenarios. Higher values indicate greater effectiveness of adversarial perturbations.}
\vspace{-5mm}
\label{accuracydrop}
\end{figure*}

\begin{figure*}[t]
\centerline{\includegraphics[width=18cm]{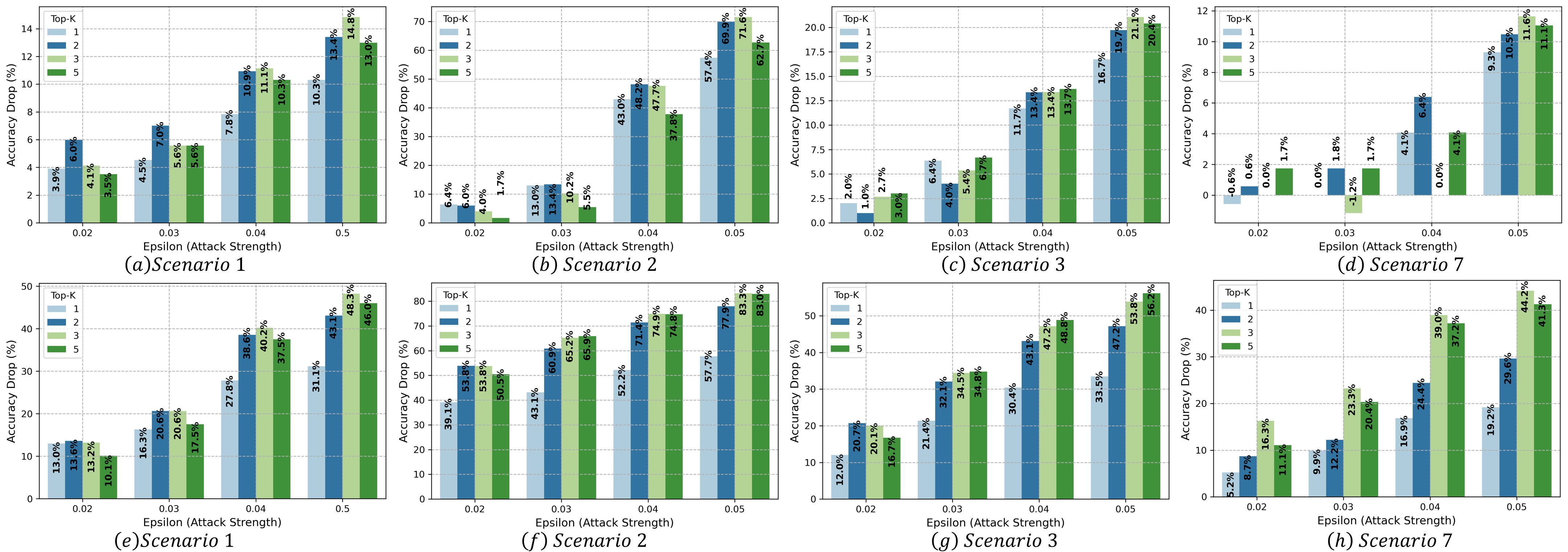}}
\caption{Raw Top-K accuracy degradation under increasing adversarial perturbation strength (\(\epsilon\)) for ResNet-50 (top) and MobileNetV2 (bottom).}
\label{accuracy_drop_trend}
\vspace{-4mm}
\end{figure*}

\subsubsection{Attack Severity Analysis}
\label{subsec:attack_severity_analysis}
This subsection presents a comparative evaluation of Gaussian noise and adversarial attacks under equivalent perturbation budgets, supported by two complementary visualizations (Figs. ~\ref{accuracydrop} and \ref{accuracy_drop_trend}). Fig.~\ref{accuracydrop} illustrates eight heatmaps comparing the impact of Gaussian noise and adversarial attacks across four different scenarios for ResNet-50 \((\text{Figs.~\ref{accuracydrop}(a)-(d)})\) and MobileNetV2 \((\text{Figs.~\ref{accuracydrop}(e)-(h)})\). Each cell denotes the difference in Top-K accuracy drop between Gaussian and adversarial settings at comparable perturbation magnitudes. Larger values signify stronger adversarial effectiveness.
We evaluate adversarial robustness by comparing adversarial perturbations with Gaussian noise of similar perceptual intensity. Specifically, we use adversarial perturbation magnitudes (\(\epsilon\)) from the set \(\{0.02, 0.03, 0.04, 0.05\}\), and Gaussian noise standard deviations (\(\sigma\)) from \(\{0.01, 0.015, 0.02, 0.025\}\). To ensure a fair comparison, we align the average per-pixel distortion by approximating \(\epsilon \approx 2\sigma\). This approximation follows from the empirical rule in Gaussian distributions, where 95\% of values lie within \(\pm2\sigma\). For example, a noise level of \(\sigma=0.025\) implies that 95\% of pixel perturbations fall within \(\pm0.05\), closely matching the adversarial bound of \(\epsilon=0.05\). Assuming perturbations follow \(\eta \sim \mathcal{N}(0, \sigma^2)\), this setup enables a direct comparison between random and structured (adversarial) perturbations in terms of distortion magnitude\cite{Gaussian_noise}.
\begin{table*}[htp!]
\centering
\small
\setlength{\tabcolsep}{3pt}
\renewcommand{\arraystretch}{0.9}

\caption{Full comparison of Random Noise ($\sigma$) vs. Adversarial Attack ($\epsilon$), across all scenarios and Top-K. Columns show (FRM+R / BR / DR / FRM+M / BM / DM) for ResNet and MobileNet side by side.}
\label{tab:all_scenarios_full}

\begin{tabular}{c c c r r r r r r c r r r r r r}
\toprule
\multirow{2}{*}{\textbf{Scenario}} & \multirow{2}{*}{\textbf{Top-K}} & \multirow{2}{*}{$\sigma$} & \multicolumn{6}{c}{\textbf{Random Noise (Accuracy)}} & \multirow{2}{*}{$\epsilon$} & \multicolumn{6}{c}{\textbf{Adversarial (Accuracy)}} \\
\cmidrule(lr){4-9} \cmidrule(lr){11-16}
 &  &  & \textbf{FRM+R} & \textbf{BR} & \textbf{DR} & \textbf{FRM+M} & \textbf{BM} & \textbf{DM} &  & \textbf{FRM+R} & \textbf{BR} & \textbf{DR} & \textbf{FRM+M} & \textbf{BM} & \textbf{DM} \\
\midrule
1 & 1 & 0.01 & \textbf{0.5979} & 0.5505 & 0.5298 & \textbf{0.5835} & 0.5381 & 0.503   & 0.02 & \textbf{0.5876} & 0.5092 & 0.3752 & \textbf{0.5835} & 0.4144 & 0.2762 \\
1 & 2 & 0.01 & \textbf{0.8288} & 0.8103 & 0.7814 & \textbf{0.8432} & 0.7773 & 0.7319  & 0.02 & \textbf{0.8144} & 0.7422 & 0.6082 & \textbf{0.8185} & 0.6432 & 0.4618 \\
1 & 3 & 0.01 & \textbf{0.9134} & 0.8824 & 0.903  & \textbf{0.9237} & 0.8969 & 0.8556  & 0.02 & \textbf{0.9134} & 0.8432 & 0.7567 & \textbf{0.8948} & 0.769   & 0.5896 \\
1 & 5 & 0.01 & \textbf{0.9814} & 0.967  & 0.9711 & \textbf{0.9876} & 0.9731 & 0.9587  & 0.02 & \textbf{0.9711} & 0.936  & 0.8639 & \textbf{0.9752} & 0.868   & 0.7092 \\
1 & 1 & 0.015 & \textbf{0.602}  & 0.5567 & 0.5381 & \textbf{0.5773} & 0.5154 & 0.4721  & 0.03 & \textbf{0.5773} & 0.503  & 0.3319 & \textbf{0.5731} & 0.3814 & 0.2020 \\
1 & 2 & 0.015 & \textbf{0.8288} & 0.7979 & 0.7752 & \textbf{0.835}  & 0.7731 & 0.6804  & 0.03 & \textbf{0.802}  & 0.7319 & 0.5546 & \textbf{0.8164} & 0.5731 & 0.3463 \\
1 & 3 & 0.015 & \textbf{0.9175} & 0.8865 & 0.9092 & \textbf{0.9134} & 0.8824 & 0.8082  & 0.03 & \textbf{0.8907} & 0.8288 & 0.6742 & \textbf{0.8927} & 0.6948 & 0.4350 \\
1 & 5 & 0.015 & \textbf{0.9793} & 0.9649 & 0.9711 & \textbf{0.9793} & 0.9731 & 0.9381  & 0.03 & \textbf{0.9484} & 0.9154 & 0.8061 & \textbf{0.9690}  & 0.7938 & 0.5381 \\
1 & 1 & 0.02  & \textbf{0.602}  & 0.5567 & 0.534  & \textbf{0.5979} & 0.503  & 0.4371  & 0.04 & 0.4453 & \textbf{0.4701} & 0.2577 & \textbf{0.4948} & 0.2659 & 0.1216 \\
1 & 2 & 0.02  & \textbf{0.835}  & 0.7979 & 0.7628 & \textbf{0.8391} & 0.7505 & 0.6309  & 0.04 & 0.6659 & \textbf{0.6927} & 0.4144 & \textbf{0.7051} & 0.3938 & 0.2041 \\
1 & 3 & 0.02  & \textbf{0.9154} & 0.8783 & 0.9072 & \textbf{0.9175} & 0.8762 & 0.7525  & 0.04 & 0.7587 & \textbf{0.7731} & 0.5340 & \textbf{0.8103} & 0.4989 & 0.2680 \\
1 & 5 & 0.02  & \textbf{0.9793} & 0.9628 & 0.9649 & \textbf{0.9773} & 0.9628 & 0.8969  & 0.04 & 0.8391 & \textbf{0.8680}  & 0.6680 & \textbf{0.8948} & 0.5938 & 0.3278 \\
\midrule
2 & 1 & 0.01 & \textbf{0.7006} & 0.6605 & 0.6655 & \textbf{0.6588} & 0.5953 & 0.5033  & 0.02 & \textbf{0.6488} & 0.5969 & 0.2357 & 0.1538 & \textbf{0.4899} & 0.0568 \\
2 & 2 & 0.01 & \textbf{0.8879} & 0.8695 & 0.8829 & \textbf{0.8695} & 0.7725 & 0.7056  & 0.02 & \textbf{0.8628} & 0.8110 & 0.3210 & 0.2959 & \textbf{0.6806} & 0.0819 \\
2 & 3 & 0.01 & \textbf{0.9498} & 0.9615 & 0.9531 & \textbf{0.9297} & 0.8645 & 0.8260  & 0.02 & \textbf{0.9297} & 0.9113 & 0.3913 & 0.4046 & \textbf{0.8010} & 0.1270 \\
2 & 5 & 0.01 & \textbf{0.9882} & 0.9799 & 0.9782 & \textbf{0.9816} & 0.9180 & 0.9214  & 0.02 & \textbf{0.9715} & 0.9598 & 0.4698 & 0.5334 & \textbf{0.8795} & 0.1906 \\
2 & 1 & 0.015 & \textbf{0.6956} & 0.6655 & 0.6538 & \textbf{0.6170} & 0.4180 & 0.3244  & 0.03 & \textbf{0.5702} & 0.5301 & 0.1956 & 0.0986 & \textbf{0.3060} & 0.0484 \\
2 & 2 & 0.015 & \textbf{0.8846} & 0.8695 & 0.8762 & \textbf{0.8160} & 0.5334 & 0.4799  & 0.03 & \textbf{0.7675} & 0.7374 & 0.2508 & 0.2023 & \textbf{0.4816} & 0.0652 \\
2 & 3 & 0.015 & 0.9448 & \textbf{0.9598} & 0.9481 & \textbf{0.9030} & 0.6438 & 0.5903  & 0.03 & 0.8461 & \textbf{0.8494} & 0.2775 & 0.2809 & \textbf{0.6053} & 0.0819 \\
2 & 5 & 0.015 & \textbf{0.9866} & 0.9765 & 0.9749 & \textbf{0.9531} & 0.7441 & 0.7056  & 0.03 & \textbf{0.9264} & 0.9214 & 0.3160 & 0.3862 & \textbf{0.7324} & 0.1204 \\
2 & 1 & 0.02  & \textbf{0.6789} & 0.6421 & 0.6454 & \textbf{0.5635} & 0.2391 & 0.2140  & 0.04 & 0.1906 & \textbf{0.2307} & 0.1053 & 0.0869 & \textbf{0.1270} & 0.0484 \\
2 & 2 & 0.02  & \textbf{0.8812} & 0.8494 & 0.8712 & \textbf{0.7391} & 0.3026 & 0.3043  & 0.04 & 0.2993 & \textbf{0.3896} & 0.1454 & 0.1404 & \textbf{0.2491} & 0.0668 \\
2 & 3 & 0.02  & \textbf{0.9431} & 0.943 & 0.9397 & \textbf{0.8428} & 0.3795 & 0.3662  & 0.04 & 0.3779 & \textbf{0.4749} & 0.1806 & 0.1923 & \textbf{0.3143} & 0.0785 \\
2 & 5 & 0.02  & \textbf{0.9849} & 0.9715 & 0.9715 & \textbf{0.9113} & 0.4832 & 0.4381  & 0.04 & 0.5418 & \textbf{0.5986} & 0.2274 & 0.2508 & \textbf{0.4163} & 0.0953 \\
\midrule
3 & 1 & 0.01 & \textbf{0.5484} & 0.5250 & 0.4983 & \textbf{0.5384} & 0.4682 & 0.3076  & 0.02 & \textbf{0.5351} & 0.4916 & 0.3846 & \textbf{0.4749} & 0.2575 & 0.1438 \\
3 & 2 & 0.01 & \textbf{0.7692} & 0.7090 & 0.6923 & \textbf{0.7157} & 0.6789 & 0.4347  & 0.02 & \textbf{0.7591} & 0.6956 & 0.5050 & \textbf{0.6254} & 0.4013 & 0.2073 \\
3 & 3 & 0.01 & \textbf{0.8862} & 0.8193 & 0.8127 & \textbf{0.8327} & 0.8193 & 0.5418  & 0.02 & \textbf{0.8628} & 0.7859 & 0.6287 & \textbf{0.7257} & 0.4715 & 0.2608 \\
3 & 5 & 0.01 & \textbf{0.9632} & 0.9230 & 0.8963 & \textbf{0.9264} & 0.8929 & 0.6789  & 0.02 & \textbf{0.9565} & 0.8996 & 0.7625 & \textbf{0.8127} & 0.6153 & 0.3745 \\
3 & 1 & 0.015 & \textbf{0.5551} & 0.5351 & 0.4816 & \textbf{0.5451} & 0.4715 & 0.2374  & 0.03 & \textbf{0.5117} & 0.4481 & 0.2909 & \textbf{0.3444} & 0.1973 & 0.1170 \\
3 & 2 & 0.015 & \textbf{0.7692} & 0.7123 & 0.6822 & \textbf{0.7123} & 0.6387 & 0.3779  & 0.03 & \textbf{0.7357} & 0.6655 & 0.3913 &\textbf{0.5016} & 0.2876 & 0.1471 \\
3 & 3 & 0.015 & \textbf{0.8762} & 0.8160 & 0.8060 & \textbf{0.8093} & 0.7558 & 0.4849  & 0.03 & \textbf{0.8494} & 0.7591 & 0.4849 & \textbf{0.5986} & 0.3578 & 0.2006 \\
3 & 5 & 0.015 & \textbf{0.9598} & 0.9163 & 0.8963 & \textbf{0.9163} & 0.8528 & 0.5986  & 0.03 & \textbf{0.9364} & 0.8628 & 0.5819 & \textbf{0.7157} & 0.4849 & 0.3110 \\
3 & 1 & 0.02  & \textbf{0.5317} & 0.5317 & 0.4381 & \textbf{0.5016} & 0.4481 & 0.1906  & 0.04 & \textbf{0.4615} & 0.3946 & 0.2006 & \textbf{0.2642} & 0.1404 & 0.1036 \\
3 & 2 & 0.02  & \textbf{0.7625} & 0.7190 & 0.4782 & \textbf{0.6521} & 0.5986 & 0.2709  & 0.04 & \textbf{0.6588} & 0.5719 & 0.2809 & \textbf{0.3678} & 0.2240 & 0.1237 \\
3 & 3 & 0.02  & \textbf{0.8729} & 0.8060 & 0.5919 & \textbf{0.7558} & 0.7157 & 0.3712  & 0.04 & \textbf{0.7725} & 0.6789 & 0.3578 & \textbf{0.4314} & 0.2909 & 0.1705 \\
3 & 5 & 0.02  & \textbf{0.9498} & 0.9096 & 0.6989 & \textbf{0.8595} & 0.8093 & 0.5117  & 0.04 & \textbf{0.8662} & 0.7926 & 0.4414 & \textbf{0.5217} & 0.3578 & 0.2600 \\
\midrule
7 & 1 & 0.01 & \textbf{0.4244} & \textbf{0.4244} & 0.3779 & \textbf{0.3895} & 0.3313 & 0.3081  & 0.02 & \textbf{0.4593} & 0.4244 & 0.2500 & \textbf{0.3488} & 0.2848 & 0.1511 \\
7 & 2 & 0.01 & \textbf{0.6453} & 0.6162 & 0.6279 & \textbf{0.5930} & 0.5348 & 0.4767  & 0.02 & \textbf{0.6395} & 0.6279 & 0.4244 & \textbf{0.6220} & 0.4360 & 0.2383 \\
7 & 3 & 0.01 & \textbf{0.7848} & 0.7325 & 0.7500 & \textbf{0.7674} & 0.7034 & 0.5988  & 0.02 & \textbf{0.7906} & 0.7325 & 0.6279 & \textbf{0.7500} & 0.5697 & 0.3139 \\
7 & 5 & 0.01 & 0.8720 & \textbf{0.8837} & 0.8779 & \textbf{0.8779} & 0.8372 & 0.7500  & 0.02 & \textbf{0.8895} & 0.8604 & 0.7441 & \textbf{0.8604} & 0.7383 & 0.4534 \\
7 & 1 & 0.015 & \textbf{0.4302} & 0.4244 & 0.3895 & \textbf{0.3662} & 0.3255 & 0.3081  & 0.03 & \textbf{0.4302} & 0.4186 & 0.1976 & \textbf{0.3313} & 0.2383 & 0.1279 \\
7 & 2 & 0.015 & 0.6511 & 0.6279 & \textbf{0.6569} & \textbf{0.5930} & 0.5232 & 0.5000  & 0.03 & \textbf{0.6511} & 0.6162 & 0.4011 & \textbf{0.5523} & 0.4011 & 0.2151 \\
7 & 3 & 0.015 & \textbf{0.7906} & 0.7383 & 0.7441 & \textbf{0.7616} & 0.6802 & 0.5872  & 0.03 & \textbf{0.7500} & 0.7441 & 0.5000 & \textbf{0.7325} & 0.5697 & 0.2732 \\
7 & 5 & 0.015 & \textbf{0.8720} & 0.8662 & 0.8895 & \textbf{0.8779} & 0.8255 & 0.7267  & 0.03 & \textbf{0.8779} & 0.8604 & 0.6453 & \textbf{0.8546} & 0.7383 & 0.3604 \\
7 & 1 & 0.02  & \textbf{0.4244} & \textbf{0.4244} & 0.3662 & \textbf{0.3546} & 0.2965 & 0.2965  & 0.04 & \textbf{0.4127} & 0.3779 & 0.1918 & \textbf{0.3023} & 0.1686 & 0.1162 \\
7 & 2 & 0.02  & \textbf{0.6511} & 0.6337 & 0.6395 & \textbf{0.5988} & 0.5116 & 0.4767  & 0.04 & \textbf{0.6279} & 0.5697 & 0.2965 & \textbf{0.5465} & 0.2790 & 0.2151 \\
7 & 3 & 0.02  & \textbf{0.7906} & 0.7616 & 0.7383 & \textbf{0.7616} & 0.6569 & 0.5755  & 0.04 & \textbf{0.7441} & 0.7325 & 0.3895 & \textbf{0.6802} & 0.3430 & 0.2558 \\
7 & 5 & 0.02  & 0.8779 & 0.8662 & \textbf{0.8953} & \textbf{0.8895} & 0.8023 & 0.6918  & 0.04 & 0.8313 & \textbf{0.8372} & 0.4767 & \textbf{0.8488} & 0.4862 & 0.3255 \\
\bottomrule
\end{tabular}
\end{table*}
From Fig.~\ref{accuracydrop}, it can be observed that for{ResNet-50} in {Scenario~2} (Fig.~\ref{accuracydrop}(b)), the model shows the highest vulnerability, with accuracy differences reaching 68.90\% (Top-3) and 66.55\% (Top-2) at $\epsilon = 0.05$.
Even under low perturbation (\(\epsilon=0.02\)), the gap already exceeds 40\%, indicating that adversarial perturbations exploit fragile decision boundaries in this setup. This could be due to cluttered or complex scenes in Scenario~2, where subtle perturbations cause the model to shift its prediction drastically. {Scenario~1} ({Fig.~\ref{accuracydrop}(a)}) also shows consistent adversarial dominance, though the magnitude is more moderate—peaking at 13.41\% in Top-3 for \(\epsilon=0.05\). {Scenario~3} ({Fig.~\ref{accuracydrop}(c)}) reflects moderate degradation, with the Top-2 accuracy difference reaching 21.74\%. These results suggest that as visual complexity increases, ResNet-50 becomes increasingly susceptible to structured perturbations.

Interestingly, in {Scenario~7} ({Fig.~\ref{accuracydrop}(d)}), the adversarial effectiveness is relatively low at smaller perturbation levels. In many of these test samples, the image contains only a single vehicle and a simple background. Such minimalistic compositions make it harder for weaker adversarial attacks (e.g., \(\epsilon=0.02\)) to significantly shift the model’s output. This leads to minimal difference between noise and adversarial cases, with even negative values in some Top-K categories. However, as the attack strength increases, the model eventually succumbs—Top-3 difference rises to 12.79\% at \(\epsilon=0.05\), showing that even simplistic scenes can be vulnerable when attacked with sufficient intensity. For {MobileNetV2}, trends vary. \textbf ({Fig.~\ref{accuracydrop}(e)}) shows severe degradation, with differences exceeding 43\% in Top-3 and Top-5 under higher \(\epsilon\), suggesting that MobileNetV2’s lighter architecture is highly sensitive to crafted perturbations in that scenario. {Scenario~3} ({Fig.~\ref{accuracydrop}(g)}) follows a similar pattern, peaking at 37.80\% Top-5 difference, again pointing to MobileNetV2's weaker feature robustness.
\begin{figure*}[t]
\centerline{\includegraphics[width=16cm]{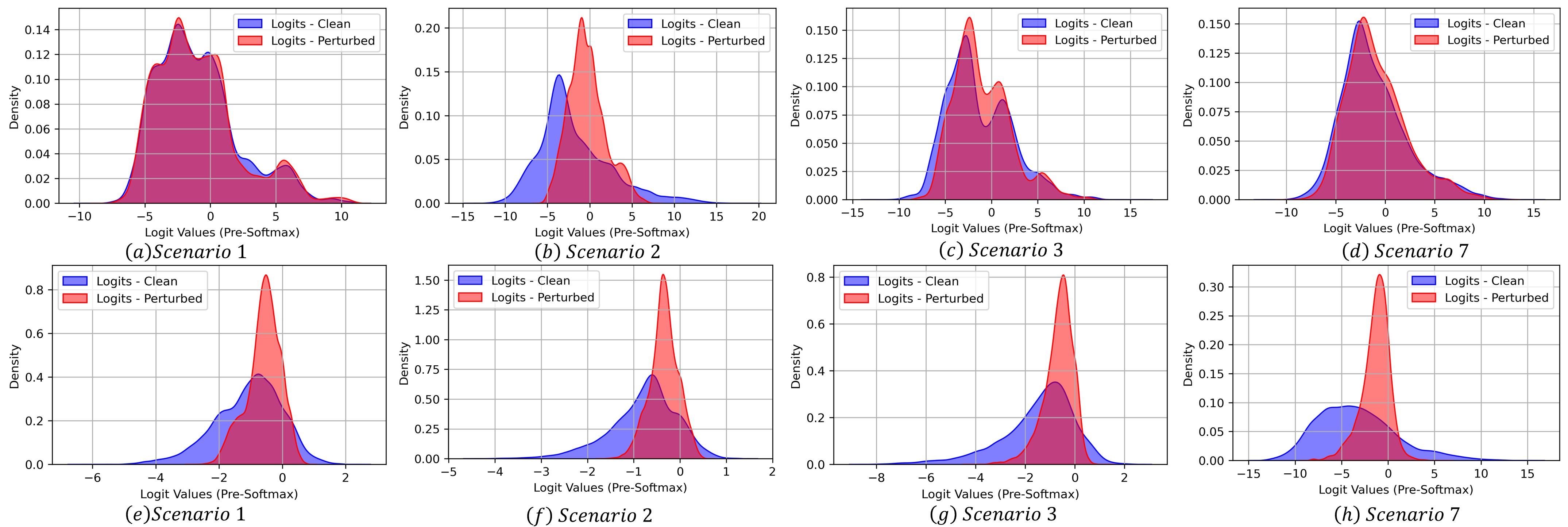}}
\vspace{-2mm}
\caption{Logit distributions (pre-softmax) for clean vs. perturbed inputs across Scenarios 1, 2, 3, and 7. Top: FRM models show overlap and sharper boundaries. Bottom: DD models exhibit compression and shifts toward zero under perturbation.}
\vspace{-5mm}
\label{fig:logit_shift}
\end{figure*}

Empirical findings, as visualized in Fig.~\ref{accuracy_drop_trend}, reaffirm and expand upon the trends observed in the comparative adversarial-versus-noise analysis in Fig.~\ref{accuracydrop}. The results in Fig.~\ref{accuracy_drop_trend} illustrate the raw accuracy degradation under increasing adversarial attack strength across four distinct scenarios for both ResNet-50 and MobileNetV2. From the figure, it is evident that the effect is most striking in Scenario~2, where both models exhibit extreme sensitivity to adversarial perturbations. MobileNetV2, in particular, shows Top-3 and Top-5 accuracy drops exceeding 74\% and 83\%, respectively. ResNet-50 also experiences significant degradation, with differences ranging from 68\% to 71\%. Similar trends can be observed throughout the figure. These results clearly demonstrate that as the perturbation magnitude (\(\epsilon\)) increases, the beam selection accuracy steadily declines in both models. However, the extent and rate of this degradation are highly dependent on the model architecture and the specific characteristics of the scene.

Collectively, the results from both Fig.~\ref{accuracydrop} and Fig.~\ref{accuracy_drop_trend} confirm that adversarial perturbations, especially those disrupting spatial relationships, pose a greater threat to model reliability. The impact varies based on scene characteristics and model architecture. While ResNet-50 shows moderate robustness due to its depth and redundancy, MobileNetV2 remains vulnerable across all settings. This is primarily due to its lightweight design, which employs depthwise separable convolutions and has fewer parameters, limiting its capacity to capture redundant or abstract features. Consequently, even small spatial shifts can significantly disrupt its specialized feature maps. In contrast, ResNet-50’s deeper architecture and residual connections facilitate more robust feature hierarchies, allowing it to better tolerate perturbations and maintain a higher level of baseline accuracy.

\subsubsection{Impact of FRM on Adversarial and Noise Attacks}
Table~\ref{tab:all_scenarios_full} presents a comprehensive comparison of model robustness under both random Gaussian noise ($\sigma$) and adversarial perturbations ($\epsilon$) across multiple scenarios and Top-K values. The models evaluated include the baseline ResNet-50 (BR), distilled ResNet (DR), FRM-enhanced ResNet (FRM+R), and their counterparts based on MobileNetV2 (BM, DM, FRM+M).

Under random noise, the baseline models experience noticeable degradation in accuracy as $\sigma$ increases. For example, in Scenario~1 with $\sigma=0.02$ at Top-1, BR and BM yield 55.67\% and 50.30\% accuracy respectively, while FRM+R and FRM+M outperform them with 60.20\% and 59.79\%. This demonstrates the effectiveness of the Feature Refining Module in mitigating random perturbations. In general, MobileNetV2-based models are more sensitive to noise due to their lightweight architecture as described earlier, but the addition of FRM noticeably improves their stability across multiple scenarios. Interestingly, the results also reveal that in some cases, introducing mild noise improves accuracy over the clean setting. For example, in Scenario~7 at Top-1, the clean accuracy of FRM+R is 43.60\%, which rises to 44.02\% at $\sigma=0.015$. Similar patterns are observed in other settings, indicating that small noise may act as a form of regularization, helping the model attend to more robust and generalizable features. This synergizes well with FRM, which likely amplifies meaningful components while reducing redundancy.

However, despite its overall benefits, there are certain cases where FRM fails to outperform its baseline. For instance, in Scenario~3 at $\epsilon=0.04$ and Top-3, FRM+R achieves 77.25\%, which is slightly lower than BR’s 80.60\%. This suggests that when adversarial noise overlaps with salient features, FRM’s refinement strategy may mistakenly suppress useful information. Another example appears in Scenario~1 at $\epsilon=0.02$ and Top-5, where BM reaches 86.80\%, just above FRM+M’s 86.48\%. Such edge cases indicate that the FRM mechanism may sometimes eliminate spatial cues that are beneficial for certain predictions.

Additionally, under adversarial conditions, all models undergo sharper performance degradation compared to random noise. For example, in Scenario~2 with $\epsilon=0.04$ at Top-1, BM drops to 12.70\%, while FRM+M recovers some accuracy and reaches 19.06\%. However, the most severe decline is observed in distilled models (DR and DM), which consistently underperform across all scenarios. In the same setting, DM achieves only 4.84\%, suggesting that its softened decision boundaries are inadequate for preserving the spatial precision required in beam selection tasks.

This weakness is primarily due to the nature of the beam prediction problem itself, where the decision often depends on a very localized spatial region of the input image, typically a small area where the user's car or device is present in the camera’s field of view. Distillation techniques tend to smooth out class boundaries and distribute learning across a wider feature space. While this is helpful for generalization in typical classification tasks, it becomes counterproductive when only a narrow region contains the signal relevant to the task. The smoothed representations learned by distilled models may dilute or suppress this fine-grained spatial information, resulting in poor robustness, especially under adversarial perturbations that exploit such localization sensitivity.

This is further illustrated by the logit distribution plots in Fig.~\ref{fig:logit_shift}, which show the changes in pre-softmax activations for clean versus perturbed inputs. The top row corresponds to the FRM-enhanced models, while the bottom row shows the same for distilled models across Scenarios 1, 2, 3, and 7. Notably, in FRM models, the distributions for clean and perturbed inputs show higher overlap and maintain sharper, more discriminative boundaries. In contrast, the DD models exhibit significant logit compression and a shift toward zero-centered values under perturbation, especially in Scenarios 2 and 3 (Figs.~\ref{fig:logit_shift}f and \ref{fig:logit_shift}g), indicating a loss of confidence and class separability. The high-density peaks around zero suggest that adversarial noise effectively collapses the model's certainty, resulting in misclassifications even under small perturbations.

In summary, the proposed FRM enhances robustness across both random and adversarial noise for the majority of cases. It especially benefits MobileNet-based architectures, which are typically more vulnerable. Nonetheless, occasional degradations highlight the need for more adaptive refinement strategies that can dynamically distinguish between noise and task-relevant features. Future work could explore attention-based or uncertainty-aware refinement mechanisms to further improve selective feature suppression and enhance generalization under challenging conditions.
\vspace{-3mm}
\section{Conclusion}
\label{conclusion}
In this work, we systematically analyzed the security vulnerabilities of vision-based mmWave beam prediction models and proposed a robust framework to enhance their adversarial robustness. Our study demonstrated that traditional white-box attack methodologies are impractical in this setting due to the inherent challenges of obtaining beam index labels. Instead, we introduced a novel black-box adversarial attack strategy that exploits spatial relationships between beam indices and user positions to generate effective perturbations, highlighting critical security concerns in vision-based beam selection.
To mitigate these vulnerabilities, we formulated an optimization framework that jointly enhances beam selection accuracy under clean conditions while improving adversarial robustness. Our proposed hybrid deep learning model, equipped with a FRM, effectively mitigates the impact of adversarial perturbations by filtering out irrelevant or misleading features. Experimental results on standard backbone architectures, including ResNet-50 and MobileNetV2, demonstrated significant improvements in both clean and adversarial accuracy, ensuring reliable beam selection in dynamic environments. Evaluations with standard backbone models, including ResNet-50 and MobileNetV2, demonstrate that the proposed method significantly enhances performance, achieving up to +21.07\% improvement in Top-K accuracy under clean conditions and up to +41.31\% gain in Top-1 robustness under adversarial attacks compared to baseline models.

\vspace{-3mm}
\bibliographystyle{IEEEtran}
\bibliography{references}

\vspace{12pt}
\end{document}